\pdfoutput=1
\documentclass[11pt,a4paper]{article}

\usepackage[utf8]{inputenc}
\usepackage[T1]{fontenc}
\usepackage{lmodern}
\usepackage[margin=1in]{geometry}
\usepackage[protrusion=true,expansion=false]{microtype}
\usepackage{ragged2e}

\usepackage{amsmath,amssymb}
\usepackage{mathtools}

\usepackage{graphicx}
\usepackage{booktabs}
\usepackage{multirow}
\usepackage{longtable}
\usepackage{array}
\usepackage{caption}
\usepackage{subcaption}
\usepackage{float}
\usepackage{placeins}
\usepackage{dblfloatfix}

\usepackage{url}
\usepackage{xspace}
\usepackage[numbers,sort&compress]{natbib}
\usepackage{hyperref}
\usepackage{cleveref}
\usepackage{authblk}


\hypersetup{
  colorlinks = true,
  urlcolor   = blue,
  linkcolor  = blue,
  citecolor  = blue,
  breaklinks = true,
  pdftitle  = {Large Language Models as Optimization Controllers: Adaptive Continuation for SIMP Topology Optimization},
  pdfauthor = {Shaoliang Yang, Jun Wang, Yunsheng Wang},
  pdfkeywords = {Topology optimization, SIMP continuation, Large language models, Online parameter control, Heaviside projection, Meta-optimization}
}

\newcommand{\rhobar}{\bar{\rho}}
\newcommand{\rhotil}{\tilde{\rho}}
\newcommand{\penal}{p}
\newcommand{\betaH}{\beta}
\newcommand{\rmin}{r_{\min}}
\newcommand{\dmove}{\delta}
\newcommand{\vf}{V_f}
\newcommand{\Gray}{\mathcal{G}}
\newcommand{\Comp}{C}
\newcommand{\Emod}{E}
\newcommand{\SIMP}{SIMP}
\newcommand{\LLM}{LLM}
\newcommand{\DNC}{DNC}
\newcommand{\OC}{OC}
\newcommand{\AMG}{AMG}
\newcommand{\MBB}{MBB}

\newcommand{\eg}{{\em e.g.}}

\title{\textbf{Large Language Models as Optimization Controllers: Adaptive Continuation for SIMP Topology Optimization}}

\author[1]{Shaoliang Yang}
\author[1]{Jun Wang\thanks{Corresponding author. E-mail: jwang22@scu.edu}}
\author[1]{Yunsheng Wang}
\affil[1]{Department of Mechanical Engineering, Santa Clara University, Santa Clara, CA 95053, USA}

\date{}

\begin{document}
\maketitle

\begin{abstract}
We present a framework in which a large language model (LLM) acts as an
\emph{online adaptive controller} for SIMP topology optimization, replacing
conventional fixed-schedule continuation with real-time, state-conditioned
parameter decisions.
At every $k$-th iteration, the LLM receives a structured
observation---current compliance, grayness index, stagnation counter,
checkerboard measure, volume fraction, and budget consumption---and outputs
numerical values for the penalization exponent~$p$, projection
sharpness~$\beta$, filter radius~$r_{\min}$, and move limit~$\delta$ via
a Direct Numeric Control interface.
A hard grayness gate prevents premature binarization, and a meta-optimization loop uses a second LLM pass to tune the agent's call frequency and gate
threshold across runs.
We benchmark the agent against four baselines---fixed (no-continuation),
standard three-field continuation, an expert heuristic, and a schedule-only
ablation---on three 2-D problems (cantilever, MBB beam, L-bracket) at
$120\!\times\!60$ resolution and two 3-D problems (cantilever, MBB beam) at
$40\!\times\!20\!\times\!10$ resolution, all run for 300 iterations.
A standardized 40-iteration sharpening tail is applied from the best valid
snapshot so that compliance differences reflect only the exploration phase.
The LLM agent achieves the lowest final compliance on every benchmark:
$-5.7\%$ to $-18.1\%$ relative to the fixed baseline, with all solutions
fully binary.
The schedule-only ablation \emph{underperforms} the fixed baseline on two of
three problems, confirming that the LLM's real-time intervention---not the
schedule geometry---drives the gain.
Code and reproduction scripts will be released upon publication.
\end{abstract}


\justifying

\bigskip

\section{Introduction}
\label{sec:intro}

\subsection{Topology Optimization and the Continuation Problem}

Topology optimization seeks a material distribution $\rho(\mathbf{x})$ within
a prescribed design domain $\Omega$ that minimizes a structural objective
while satisfying the equilibrium and a volume constraint.
The SIMP method~\citep{Bendsoe1989,Sigmund2001} remains the dominant
computational paradigm: element densities $\rho_e \in [0,1]$ serve as
design variables, and the power-law stiffness interpolation
\begin{equation}
  \Emod(\rho_e) \;=\; \Emod_{\min} + \rho_e^{\,\penal}
  \bigl(\Emod_0 - \Emod_{\min}\bigr),
  \qquad \penal > 1,
  \label{eq:simp}
\end{equation}
penalizes intermediate densities to promote solid-void solutions.
In the basic SIMP formulation, $\rho_e$ in Eq.~\eqref{eq:simp} denotes the
raw element density; in the three-field extension introduced below, the
power law acts instead on the projected physical density~$\rhotil_e$
(see Eq.~\eqref{eq:simp:method} in Sect.~\ref{sec:formulation}).
In the three-field formulation of \citet{Wang2011}, the raw design field
$\boldsymbol{\rho}$ passes through two successive projections before entering the finite-element solve: a linear density filter of radius~$\rmin$
produces a smoothed field~$\rhobar$, which is then mapped to a physical
density~$\rhotil$ by the Heaviside projection
\begin{equation}
  \rhotil_e \;=\;
  \frac{\tanh(\betaH\,\eta) + \tanh\!\bigl(\betaH(\rhobar_e - \eta)\bigr)}
       {\tanh(\betaH\,\eta) + \tanh\!\bigl(\betaH(1-\eta)\bigr)},
  \label{eq:heaviside}
\end{equation}
where $\eta \in (0,1)$ is a threshold parameter and $\betaH \geq 1$
controls the sharpness of the projection.
As $\betaH \to \infty$, Eq.~\eqref{eq:heaviside} approaches a step function
and $\rhotil$ becomes binary; for $\betaH = 1$ the mapping is nearly
linear \citep{Wang2011,Lazarov2016}.
This decoupling of the design space from the physical density is the key advantage of the three-field approach: the filter enforces a minimum length
scale while the Heaviside projection drives binarization, and the two
processes can be controlled independently through $\rmin$ and $\betaH$,
respectively.

Despite the elegance of this formulation, practical performance is acutely
sensitive to the \emph{continuation schedule}: the trajectory along which
$\penal$, $\betaH$, $\rmin$, and the Optimality Criteria (\OC{}) move
limit~$\dmove$ are varied over the course of the solve.
Increasing $\betaH$ too early, before a coherent structural topology has
formed, locks the design into a poor local minimum from which the gradient method cannot escape; conversely, deferring sharpening wastes computational budget on an excessively gray intermediate field~\citep{Sigmund2007,Lazarov2016}.
Standard practice relies on monotone schedules calibrated by the
practitioner---typically a linear ramp for $\penal$ and geometric doubling for
$\betaH$ at fixed iteration intervals---that are tuned for a particular problem
class, mesh resolution, and iteration budget and must be re-calibrated
whenever any of these change~\citep{Lazarov2016,Wang2011}.

\subsection{Limitations of Fixed-Schedule Continuation}

The fundamental limitation of fixed-schedule continuation is that it is
\emph{open-loop}: the schedule commits to a progression of parameter values
independent of what is actually occurring in the density field.
Consider two runs initialized identically but perturbed by different random
seeds: the grayness~$\Gray$ (fraction of elements with intermediate density)
may differ substantially at any given iteration, yet both runs receive the
same $\betaH$ update regardless.
If $\Gray$ is still declining rapidly---indicating that material is actively
consolidating under penalization---a premature $\betaH$ increase interrupts this process and crystallizes a sub-optimal topology.
Conversely, if $\Gray$ has plateaued while $\penal$ is still low, the schedule may continue to invest iterations in a gray, under-penalized phase rather than advancing toward sharpening.

This sensitivity is not a minor practical inconvenience.
Our experiments confirm that a schedule-only controller that runs the same
phase structure as an adaptive agent---but without observing the actual solver
state---achieves compliance \emph{worse} than the fixed no-continuation
baseline on two of three tested problems (cantilever: $+0.43\%$;
\MBB{} beam: $+1.09\%$ versus fixed; see Table~\ref{tab:results_main}
and Sect.~\ref{sec:ablation}).
The result is striking: blindly applying a structured schedule is not merely
neutral but actively harmful, because each phase transition locks in parameter values that may be premature or delayed for the actual trajectory of the
specific run.
This motivates the need for a \emph{closed-loop}, state-conditioned controller that can observe, reason about, and react to the evolving topology
in real time.
More broadly, this limitation exemplifies a recurring tension in computational mechanics: iterative numerical methods often depend on
hyperparameter schedules that embed implicit assumptions about the solution
trajectory, yet the optimal schedule cannot be known without first solving
the problem.
Continuation in SIMP is a particularly clean instance of this tension,
because the four control parameters ($\penal$, $\betaH$, $\rmin$, $\dmove$)
have direct physical interpretations and the feedback signal (grayness,
compliance, volume) is rich and inexpensive to compute.
This makes it an ideal testbed for investigating whether
\emph{language-model-guided} closed-loop control can replace hand-tuned
open-loop schedules in physics-based solvers.

\subsection{Large Language Models as Online Optimizers}

Recent advances in large language models (\LLM{}s) have demonstrated
structured reasoning capacity well beyond pattern matching: models such as
Gemini~\citep{Team2024Gemini}, GPT-4~\citep{OpenAI2023GPT4}, and
Claude~\citep{Anthropic2024} achieve expert-level performance on mathematics,
scientific question answering, and multi-step planning tasks that require the
integration of symbolic reasoning with domain knowledge~\citep{Wei2022,Yao2023ReAct}.
Of particular relevance is the paradigm of \emph{\LLM{}-as-agent}: the
model observes a state representation, reasons about it in natural language,
and emits a structured action~\citep{Yao2023ReAct,Park2023}.
This paradigm has been applied to code generation~\citep{Chen2021Codex},
robotic control~\citep{Driess2023PaLME}, materials
discovery~\citep{Boiko2023}, and multi-objective
optimization~\citep{Liu2024LLMOptim}, but has not, to our knowledge, been
applied to the \emph{online} control of all four continuation parameters
jointly in density-based topology optimization.
Closest to our setting, \citet{RojasStolpe2015} use convergence-triggered
rules to adapt~$\penal$ alone, and \citet{AutoProjTO2025} use a fixed
grayness rule for $\betaH$ steps; neither employs a learned, state-conditioned
policy over the full parameter set.

The \LLM{}-as-optimizer framing is attractive for this application for
several reasons, and it can be situated within the Dynamic Algorithm
Configuration (DAC) paradigm~\citep{BiedenkappDAC2020,AdriaensenAutoDAC2022}: rather than learning a policy via reinforcement learning, the \LLM{} with
SIMP domain knowledge acts directly as the policy, sidestepping the large number of environment interactions that RL-based DAC requires.
Conceptually, the agent's behavior also echoes \emph{curriculum
learning}~\citep{BengioCurriculum2009}: SIMP continuation is a form
of gradually increasing difficulty, and the \LLM{} reads the optimizer's
current state to pace this curriculum adaptively rather than following a
predetermined schedule~\citep{BengioMLCO2021}.
First, it requires \emph{no training or fine-tuning}: domain knowledge about
SIMP continuation---when to ramp $\penal$, how to interpret grayness
stagnation, why premature binarization is harmful---can be encoded directly
in the system prompt.
Second, the output of the \LLM{} is immediately interpretable: it emits
$(\penal, \betaH, \rmin, \dmove)$ with a one-line rationale, providing a
natural audit trail of every parameter decision.
Third, the architecture generalizes trivially across problem geometries,
boundary conditions, and mesh resolutions without re-calibration, since the model observes the actual state of the current run rather than
pattern-matching to a stored problem type.
These properties contrast sharply with black-box surrogate or Bayesian
optimization approaches, which optimize over entire run trajectories and typically require either many historical runs or
problem-specific parameterization~\citep{Shahriari2016,Hutter2019}.

From a broader perspective, this work sits at the convergence of two
trends that have developed largely in isolation.
On one side, the computational mechanics community has accumulated decades
of practitioner knowledge about continuation strategies---knowledge that is
typically transmitted through tutorials, code comments, and trial-and-error
rather than through formal algorithms.
On the other side, the \LLM{} community has demonstrated that language
models can absorb and operationalize such informal domain knowledge when
it is encoded as natural-language instructions~\citep{Wei2022,Yao2023ReAct}.
The present work connects these two lines by showing that the practitioner's
understanding of SIMP continuation---``do not raise $\betaH$ while the
topology is still gray''---can be made into an executable, state-conditioned
policy by encoding it in a system prompt and letting the model apply it
adaptively to each run.
The result is neither a replacement for the domain expert nor a black-box surrogate, but rather a mechanism for translating qualitative engineering judgment into quantitative solver control at negligible API cost, making the approach accessible to any research group with API access.

\subsection{Contributions}
\label{sec:contributions}

This paper makes the following contributions:

\begin{enumerate}

  \item \textbf{Direct Numeric Control (DNC) architecture.}
  We design an \LLM{} controller that operates in a \emph{direct numeric}
  mode: rather than selecting a discrete phase label that maps to preset values, the model outputs floating-point values for $\penal$, $\betaH$,
  $\rmin$, and $\dmove$ at every $k$-th iteration.
  Safety rails clamp outputs to physically valid ranges; a hard grayness gate
  ($\betaH \leq 8$ while $\Gray > 0.20$) enforces the empirically-derived constraint that Heaviside sharpening degrades solution quality when substantial gray material remains.

  \item \textbf{Meta-optimization loop.}
  A second LLM pass reads post-run summary statistics and proposes updates
  to the primary agent's own hyperparameters --- grayness gate threshold, LLM
  call frequency, and phase timing constants --- via regex patching of the
  source constants.
  This creates a two-level adaptive system: the inner agent adapts parameter values within a run; the outer loop adapts the agent's own decision rules
  across runs.

  \item \textbf{Controlled comparison with standardized tail.}
  We construct a rigorous experimental protocol in which all four continuation
  controllers (three-field continuation, expert heuristic, schedule-only
  ablation, and LLM agent) share an \emph{identical} $40$-iteration sharpening
  tail ($\penal = 4.5$, $\betaH = 32$, $\rmin = 1.20$, $\dmove = 0.05$)
  applied from the best valid snapshot stored during the main loop; the fixed controller serves as a no-intervention reference without a tail.
  This controlled protocol ensures that all compliance differences among
  continuation controllers at the termination of a run are attributable
  exclusively to the quality of the topology produced during the main
  exploration phase---the domain of the controller under comparison.

  \item \textbf{Ablation confirming LLM causality.}
  The schedule-only ablation controller---which reproduces the LLM's
  four-phase schedule structure without querying the LLM API---
  isolates the contribution of online adaptation from the contribution of
  the phase structure alone.
  Its consistent underperformance relative to the LLM agent, and its
  underperformance relative to the fixed baseline on two of three benchmarks,
  constitutes strong evidence that the LLM's real-time decisions---not the
  phase structure---are the source of the observed compliance gains.

\end{enumerate}

\section{Related Work}
\label{sec:related}

This section reviews prior work in four areas that directly motivate and
contextualize the present contribution:
SIMP topology optimization and continuation strategies
(Sect.~\ref{sec:rw_simp}), machine learning applied to topology optimization
(Sect.~\ref{sec:rw_ml}), Bayesian optimization and dynamic algorithm
configuration for solver hyperparameters (Sect.~\ref{sec:rw_hpo}), and large
language models as optimization agents (Sect.~\ref{sec:rw_llm}).
Throughout, we highlight the gap that the present work fills: no prior method has used a learned agent to jointly and adaptively control all four
continuation parameters ($\penal$, $\betaH$, $\rmin$, $\dmove$) of the
three-field \SIMP{} formulation on a per-iteration basis.

\subsection{SIMP Topology Optimization and Continuation}
\label{sec:rw_simp}

Density-based topology optimization originates with the homogenization
approach of \citet{BendsoeKikuchi1988} and the closely related
\SIMP{} material interpolation was introduced independently
by \citet{BendsoeSigmund1999} and \citet{ZhouRozvany1991}.
\citet{Bendsoe1989} and \citet{Sigmund2001} established
the computational foundations and disseminated accessible reference
implementations.
The three-field formulation---design variables $\boldsymbol{\rho}$ filtered
through a density regularisation of radius $\rmin$ and then projected via
a Heaviside function of sharpness $\betaH$---was introduced
by \citet{Wang2011} following earlier filter work by \citet{Bourdin2001}
and \citet{BrunsTortorelli2001} and projection work by \citet{Guest2004}.
The Helmholtz PDE-based filter of \citet{LazarovSigmund2011} provides a
scalable implementation for large-scale problems.
Length-scale and manufacturability constraints arising in this formulation
are analysed comprehensively by \citet{Lazarov2016}.
The grayness index is used throughout the present work as a gate criterion
derives from the morphology-based filter of \citet{Sigmund2007}.

The core difficulty that motivates our approach is the sensitivity of
three-field \SIMP{} to its continuation schedule.
\citet{StoMeSvanberg2001} demonstrated theoretically that the trajectories traced by penalization continuation methods can be discontinuous, and naive schedules may fail to converge to black-and-white designs.
\citet{GuestElimBeta2011} proposed eliminating $\betaH$-continuation
altogether through formulation modifications, but this is not always
applicable.
The most directly related prior work on \emph{automatic} continuation is
\citet{RojasStolpe2015}, who propose convergence-based triggers for
automatically incrementing the penalization exponent $\penal$ in standard
\SIMP{}.
Their method adapts $\penal$ alone and uses a deterministic trigger rule;
it does not adapt $\betaH$, $\rmin$, or $\dmove$ jointly, nor does it use any learning-based controller.
A very recent independent development by \citet{AutoProjTO2025} proposes
using the grayness indicator as an automatic stopping criterion for $\betaH$
steps in three-field \SIMP{}, again via a fixed rule rather than a
state-conditioned policy.
The present work extends this line of research by replacing deterministic
trigger rules with a language model that observes the full solver state and
outputs all four continuation parameters simultaneously.

The Optimality Criteria (\OC{}) update method used throughout the paper
was formalized by
\citet{Svanberg1987}, with a more general globally convergent variant
(GCMMA) in \citet{Svanberg2002}.
For large-scale 3-D problems the \AMG{}-preconditioned solver from
\citet{Bell2023PyAMG} is used, consistent with giga-voxel implementations
such as \citet{Aage2017}.
A modern 99-line reference implementation including the three-field
formulation is given by \citet{FerrariSigmund2020}.

\subsection{Machine Learning for Topology Optimization}
\label{sec:rw_ml}

A growing body of work applies machine learning to accelerate or supplement
iterative topology optimization.
\citet{Shin2023review} surveys the field and identifies
three main paradigms: (i)~prediction of near-optimal topologies from boundary
conditions using CNNs or U-Net-like architectures~\citep{Sosnovik2019},
(ii)~surrogate models for the finite-element solve that replace or reduce
the number of FEA calls~\citep{Cang2019,Abueidda2020}, and (iii)~generative
models that sample from a learned distribution of optimal
topologies~\citep{NieTopologyGAN2021,MazeDiffusion2023}.

Neural reparameterization offers a complementary perspective:
\citet{Hoyer2019} showed that optimizing neural network weights as implicit density representations exploits inductive bias to regularize the design
space, and subsequent work has made this practically competitive with
standard \SIMP{} while retaining mesh-independence.
The self-directed online machine learning framework of \citet{DengSOLO2022}
integrates a DNN surrogate with FEM to dynamically generate training data
near the predicted optimum, reducing computational cost by orders of
magnitude.
Reinforcement learning has been applied to gradient-free topology
optimization via sequential element removal~\citep{BrownRL2022} and to
truss TO under constraints~\citep{Hayashi2020}.

A key distinction separates all of these methods from the present work.
CNN and generative approaches learn a mapping from the problem specification
to topology, effectively replacing the optimizer.
Surrogate approaches learn to approximate the forward model.
None of these methods addresses the \emph{continuation scheduling problem}:
given a running iterative \SIMP{} solver, how should the continuation
parameters be adapted in real time to prevent premature binarization and
local minima?
The present work occupies this gap, using a language model not as a
topology predictor or surrogate, but as a \emph{per-iteration parameter
controller} that observes solver state and adjusts continuation parameters
accordingly.

\subsection{Hyperparameter Optimization and Dynamic Algorithm Configuration}
\label{sec:rw_hpo}

Hyperparameter optimization (HPO) methods---including Bayesian
optimization~\citep{Shahriari2016,Snoek2012}, the Tree-structured Parzen
Estimator~\citep{Bergstra2011TPE}, SMAC~\citep{Hutter2011SMAC}, and
Hyperband~\citep{LiHyperband2018}---treat algorithm hyperparameters as
static quantities to be tuned before or between runs.
Population-Based Training~\citep{JaderbergPBT2017} extends this to learn
hyperparameter \emph{schedules} across a population of models during
training.
Comprehensive surveys of HPO methods are provided by \citet{Hutter2019}
and \citet{BischlHPO2023}.

These offline or between-run HPO methods are fundamentally different from
what is needed for \SIMP{} continuation, where the optimal parameter values
depend on the current topology---information available only during the run.
This motivates the \emph{Dynamic Algorithm Configuration} (DAC) paradigm,
formalised by \citet{BiedenkappDAC2020} and surveyed comprehensively
by \citet{AdriaensenAutoDAC2022}.
In DAC, algorithm parameters are treated as an MDP policy to be learned via
reinforcement learning, with the optimizer's current state as observations.
Concrete DAC applications include learning step-size policies for
CMA-ES~\citep{Shala2020}, and the DACBench benchmark
library~\citep{EimerDACBench2021} provides standardised evaluation
protocols.

The present work can be viewed as a \emph{non-RL DAC agent}: instead of
learning a policy via trial-and-error, an LLM with SIMP domain knowledge
embedded in its system prompt acts as the policy, observing
solver state and emitting parameter actions every $k$ iterations.
This avoids the large number of environment interactions required by RL-based
DAC methods while leveraging the pre-trained reasoning capabilities of
frontier models.
The meta-optimization outer loop---which tunes the agent's own
hyperparameters across runs using a second LLM pass---is related to
\emph{algorithm configuration}~\citep{Hutter2011SMAC} and
PBT~\citep{JaderbergPBT2017}, but operates at a higher level by adapting
the controller's decision rules rather than its policy parameters.

\subsection{Large Language Models as Optimization Agents}
\label{sec:rw_llm}

The paradigm of using large language models as agents that perceive structured observations and emit actions has developed rapidly since
the introduction of GPT-3~\citep{Brown2020} and chain-of-thought
prompting~\citep{Wei2022}.
The ReAct framework of \citet{Yao2023ReAct} showed that interleaving reasoning traces with external tool calls substantially improves performance
on multi-step decision tasks.
Toolformer~\citep{SchickToolformer2023} demonstrated that LLMs can
learn to invoke external computational tools---including calculators and
solvers---via self-supervised training, establishing the paradigm of LLM-driven
solver invocation that the present work instantiates in the \SIMP{} setting.
Embodied agents such as PaLM-E~\citep{Driess2023PaLME} and
Voyager~\citep{WangVoyager2024} have shown that LLMs can maintain and
exploit long-term context about a changing environment---a capability
directly exploited here through the agent's running observation of
grayness, compliance, and budget-consumed fraction.

LLMs have been deployed as \emph{direct optimizers} in several recent
works.
OPRO~\citep{Yang2024OPRO} treats the LLM as an optimizer over
text prompts, using previous (solution, score) pairs in the context window
as a few-shot surrogate gradient.
FunSearch~\citep{Romera2024FunSearch} combines evolutionary search with
LLM code generation to discover new mathematical functions, achieving
state-of-the-art results on the cap-set problem.
ReEvo~\citep{YeReEvo2024} uses LLM reflections as ``verbal gradients''
to evolve heuristics for combinatorial optimization, achieving competitive
results across six benchmark problem classes.
These methods treat the LLM as a batch optimizer that proposes new
solutions in context.
In contrast, the present work uses the LLM as an \emph{online
per-iteration controller}: rather than searching over a solution space, the agent observes the evolving physical state of a running solver and
outputs exact floating-point parameter values that govern the solver's
next iteration.

Iterative self-refinement frameworks such as
Self-Refine~\citep{MadaanSelfRefine2023} and
Reflexion~\citep{ShinnReflexion2023} show that LLMs improve their
outputs when given structured feedback---a mechanism that the present
system uses explicitly through the per-call compliance and grayness
observations fed back to the model.
LLM-driven scientific discovery has been demonstrated in chemistry
by \citet{Boiko2023} and in mathematics by FunSearch~\citep{Romera2024FunSearch},
while autonomous research pipelines are reviewed by \citet{WuEvoLLM2025}.
In structural and mechanical engineering, applications of LLMs remain
nascent~\citep{Rios2023}; no prior work has applied an LLM to the online control of iterative numerical solvers in topology optimization.

An important conceptual connection is to \emph{curriculum learning}
\citep{BengioCurriculum2009}, which formalizes the insight that gradual difficulty scheduling improves convergence in nonconvex problems.
\SIMP{} continuation is mathematically a form of curriculum learning:
the optimization landscape is initially smooth (low $\penal$, low $\betaH$)
and progressively made harder (binary) as the algorithm matures.
The LLM agent's behaviour---holding $\betaH \leq 8$ while $\Gray > 0.20$
and deferring sharpening by $\approx\!70$--$90$ iterations relative to
fixed-schedule controllers---can be interpreted as \emph{adaptive curriculum
pacing}: the agent reads the learner's (optimizer's) current state to decide when to advance the difficulty rather than following a predetermined pace.
This connection situates the approach within the broader literature on
learned curriculum strategies~\citep{BengioMLCO2021} and dynamic algorithm
configuration~\citep{AdriaensenAutoDAC2022}.

\section{Methodology}
\label{sec:method}

\begin{figure}[!htbp]
  \centering
  \includegraphics[width=\linewidth]{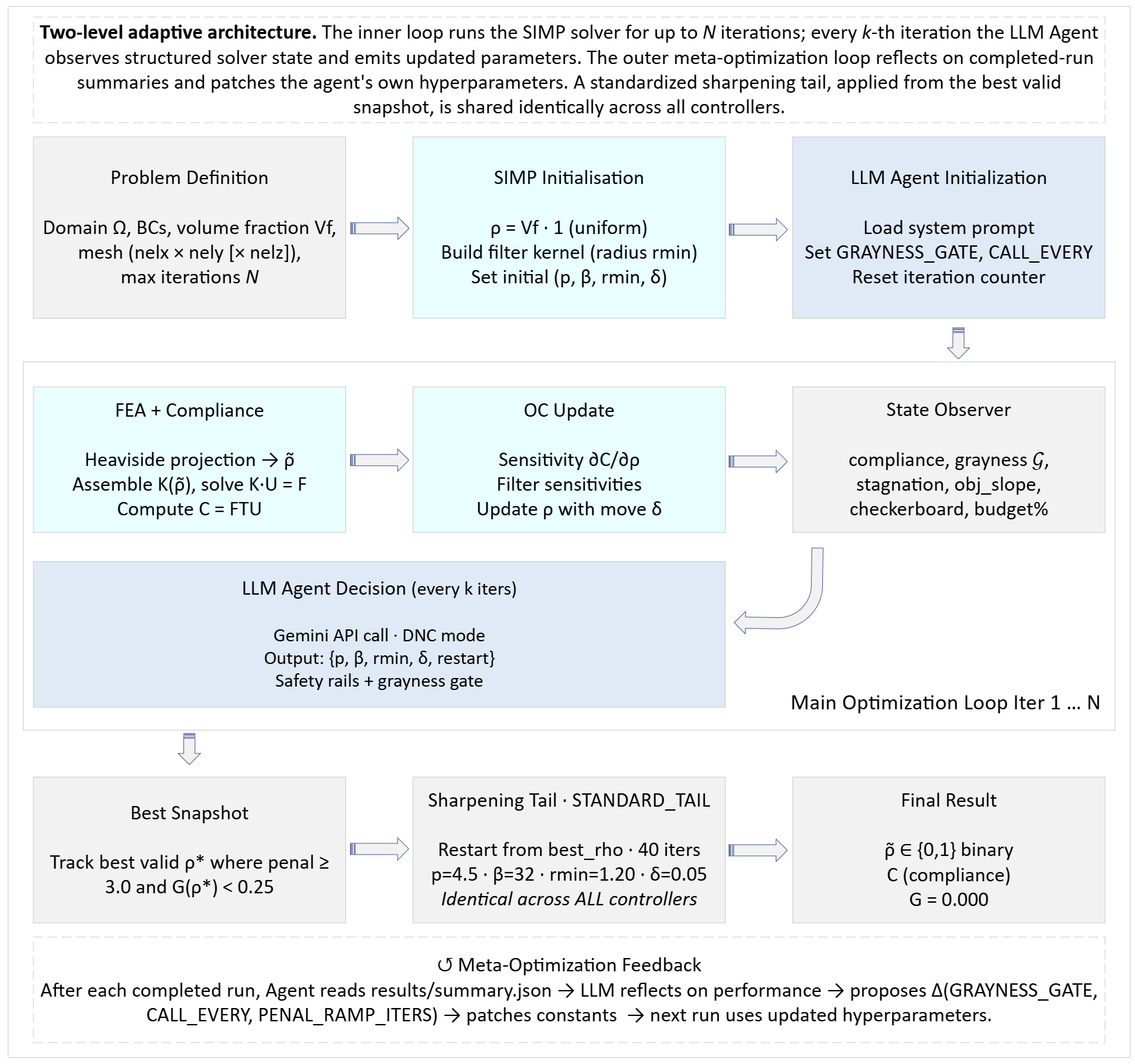}
  \caption{Overall two-level system pipeline. The inner loop executes
    the \SIMP{} solver for up to $N$ main-loop iterations; every
    $k$\textsuperscript{th} iteration, the LLM agent observes the
    structured solver state and emits updated solver parameters. The outer meta-optimization loop reflects on completed-run summaries and
    patches the agent's own hyperparameters for the next run. A
    standardized $40$-iteration sharpening tail, applied from the best
    valid intermediate snapshot, is shared identically across all four
    continuation controllers, ensuring that all compliance differences
    at run termination are attributable exclusively to main-loop behavior}
  \label{fig:pipeline}
\end{figure}

This section describes the complete two-level adaptive architecture.
Sect.~\ref{sec:formulation} establishes the three-field SIMP formulation and
the continuation parameter space.
Sect.~\ref{sec:agent} presents the LLM agent and its Direct Numeric Control
(\DNC{}) interface.
Sect.~\ref{sec:meta} describes the meta-optimization outer loop.
Sect.~\ref{sec:baselines} defines the four baseline controllers used for
comparison.
Sect.~\ref{sec:protocol} specifies the experimental protocol, including the
standardized sharpening tail that ensures fair comparison.
The overall pipeline is illustrated in Fig.~\ref{fig:pipeline}.

\subsection{Three-Field \SIMP{} Formulation}
\label{sec:formulation}

\begin{figure}[!htbp]
  \centering
  \includegraphics[width=\linewidth]{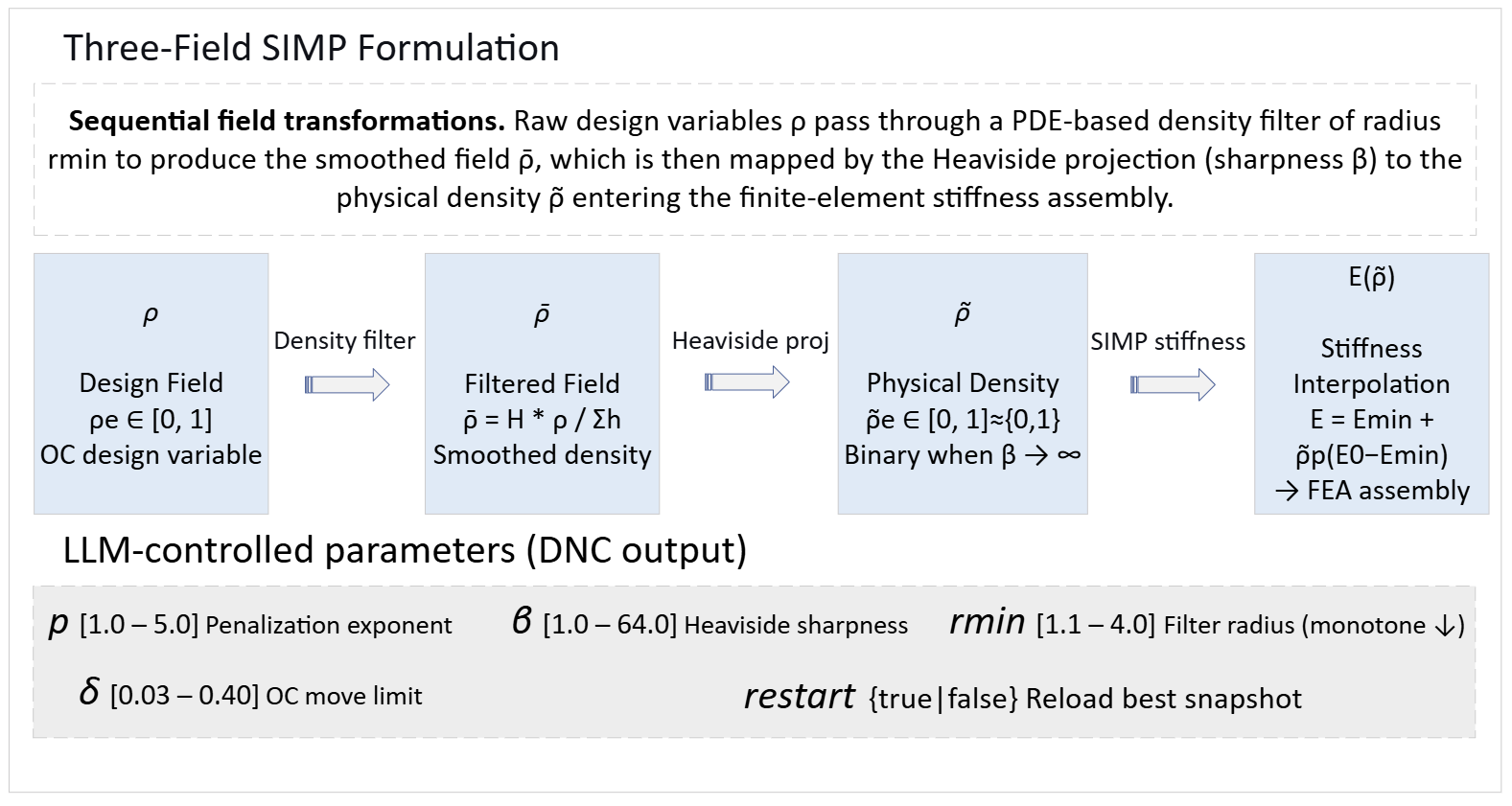}
  \caption{Three-field \SIMP{} formulation. Raw design variables
    $\rho \in [0,1]$ pass through a linear density filter of
    radius~$\rmin$ to produce a smoothed field~$\rhobar$, which is then
    mapped by the Heaviside projection (sharpness~$\betaH$,
    threshold~$\eta = 0.5$) to the physical density~$\rhotil$ entering
    stiffness assembly. The five LLM-controlled parameters
    ($\penal$, $\betaH$, $\rmin$, $\dmove$, restart flag) output by the
    Direct Numeric Control interface are listed in the lower panel with
    their admissible ranges}
  \label{fig:threefield}
\end{figure}

We adopt the three-field \SIMP{} formulation of~\citet{Wang2011}, in which
the raw design field $\boldsymbol{\rho}$ passes through two successive
transformations before entering the finite-element stiffness assembly
(Fig.~\ref{fig:threefield}).

\paragraph{Density filter.}
A spatially weighted average of radius~$\rmin$ maps the design field to the
filtered field:
\begin{equation}
  \rhobar_e
  \;=\;
  \frac{\displaystyle\sum_{i \in \mathcal{N}_e} H_{ei}\,\rho_i}
       {\displaystyle\sum_{i \in \mathcal{N}_e} H_{ei}},
  \quad
  H_{ei} = \max\!\bigl(0,\, \rmin - \|\mathbf{x}_e - \mathbf{x}_i\|\bigr),
  \label{eq:filter}
\end{equation}
where $\mathcal{N}_e$ is the neighbourhood of element~$e$ within
radius~$\rmin$.
The filter enforces a minimum length scale and eliminates checkerboard
instabilities~\citep{Bourdin2001}.

\paragraph{Heaviside projection.}
The filtered field is projected to a near-binary physical density via the
smooth Heaviside function~\citep{Wang2011,Guest2004}:
\begin{equation}
  \rhotil_e
  \;=\;
  \frac{\tanh(\betaH\,\eta)
        + \tanh\!\bigl(\betaH(\rhobar_e - \eta)\bigr)}
       {\tanh(\betaH\,\eta)
        + \tanh\!\bigl(\betaH(1-\eta)\bigr)},
  \label{eq:heaviside:method}
\end{equation}
with threshold $\eta = 0.5$.
As $\betaH \to \infty$, $\rhotil$ converges to a step function and the
design becomes fully binary; at $\betaH = 1$ the mapping is nearly linear.

\paragraph{Stiffness interpolation.}
The physical density determines element stiffness via the \SIMP{}
power law~\citep{Bendsoe1989}:
\begin{equation}
  \Emod(\rhotil_e)
  \;=\;
  \Emod_{\min}
  + \rhotil_e^{\,\penal}\bigl(\Emod_0 - \Emod_{\min}\bigr),
  \quad \penal > 1,
  \label{eq:simp:method}
\end{equation}
where $\Emod_0$ is the stiffness of the solid material and
$\Emod_{\min} = 10^{-9}\Emod_0$ prevents singularity.

\paragraph{Compliance minimization.}
The structural compliance objective is
\begin{equation}
  \Comp
  \;=\;
  \mathbf{F}^{\mathsf{T}}\mathbf{U}
  \;=\;
  \mathbf{U}^{\mathsf{T}}\mathbf{K}(\rhotil)\,\mathbf{U},
  \label{eq:compliance}
\end{equation}
where $\mathbf{K}$ is the assembled stiffness matrix and $\mathbf{U}$
satisfies $\mathbf{K}\mathbf{U} = \mathbf{F}$.
Design updates are computed by the \OC{} method with move limit~$\dmove$; no additional optimizer (MMA, etc.) is used.

\paragraph{Grayness index.}
Following \citet{Sigmund2007}, we quantify intermediate-density material
by the grayness index
\begin{equation}
  \Gray
  \;=\;
  \frac{4}{N_e}\sum_{e=1}^{N_e}\rhotil_e\!\left(1-\rhotil_e\right)
  \;\in\;[0,1],
  \label{eq:grayness}
\end{equation}
where $N_e$ is the total number of elements.
$\Gray = 0$ corresponds to a fully binary design; $\Gray = 1$ to a fully gray one.
The LLM agent observes $\Gray$ at every call and uses it to pace the
Heaviside sharpening schedule.

\subsection{LLM Agent and Direct Numeric Control Interface}
\label{sec:agent}

\begin{figure}[!htbp]
  \centering
  \includegraphics[width=\textwidth]{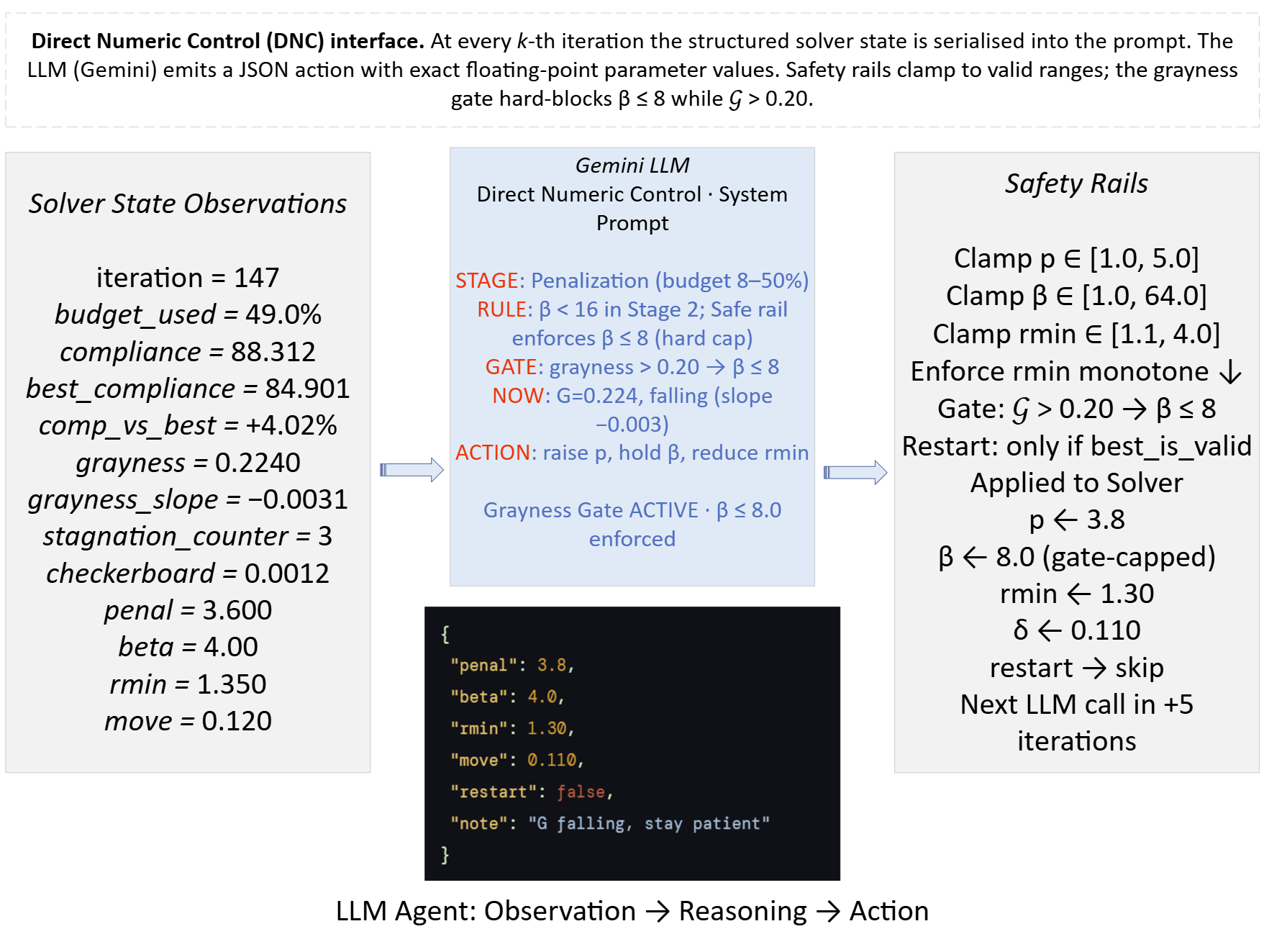}
  \caption{LLM agent decision loop at iteration~$147$ ($49\%$ of budget
    consumed). \textit{Left}---state observation: the solver's current
    state is serialized into thirteen scalar quantities. The compliance~$\Comp$
    measures global structural flexibility (lower is stiffer and better);
    $\Comp^*$ is the lowest valid compliance seen so far.
    The grayness~$\Gray \in [0,1]$ quantifies intermediate-density material:
    $\Gray = 0$ is a fully binary (solid/void) design; $\Gray > 0.20$
    indicates substantial unresolved material that would degrade
    manufacturability. The stagnation counter counts consecutive iterations
    without improvement in~$\Comp^*$; the checkerboard index detects
    non-physical alternating-density patterns.
    \textit{Centre}---reasoning: the model identifies the current budget phase
    (Stage~2, penalization), notes that the grayness gate is active
    ($\Gray = 0.224 > 0.20$), and decides to raise~$\penal$ while
    holding~$\betaH$ to avoid premature binarization.
    \textit{Right}---safety rails: the JSON output is clamped to physically
    valid ranges; the grayness gate hard-caps $\betaH \leq 8$ while
    $\Gray > 0.20$; $\rmin$ is enforced monotonically non-increasing; and
    restarts are permitted only when a valid best snapshot exists}
  \label{fig:agent}
\end{figure}

Figure~\ref{fig:agent} illustrates the per-iteration decision loop of the
LLM agent.
Every $k$\textsuperscript{th} iteration (default $k = 5$), the solver state
is serialized into a structured text prompt and submitted to the Gemini API
(model \textit{gemini-3.1-flash-lite-preview}\footnote{Model accessed in
preview form, March 2026; the exact identifier reflects the version
available at the time of experimentation and may differ from any publicly
released variant.}).
The model returns a JSON object with exact floating-point values for all four solver parameters plus a restart flag; a reasoning note is also requested for interpretability, but does not affect the solver.

\paragraph{Observation vector.}
The prompt is built from thirteen scalar quantities drawn from the state
vector at each call:
(1)~\textit{iteration indices}: iteration number, budget consumed (\% of $N$),
iterations since best;
(2)~\textit{compliance signals}: current compliance~$\Comp$, best valid
compliance~$\Comp^*$, relative deviation $(C-C^*)/C^*$, relative change
over 1 and 5 steps, and an objective slope;
(3)~\textit{topology signals}: grayness~$\Gray$, grayness-proxy slope
(computed as $s_{\Gray} = -|\Delta\Comp_5/\Comp|\cdot\mathbf{1}[\text{stag}<3]$,
where $\Delta\Comp_5$ is the relative compliance change over the last five
iterations; the compliance-based formulation captures whether the optimizer
is still making productive progress), checkerboard index, volume fraction; and
(4)~\textit{current parameters}: $\penal$, $\betaH$, $\rmin$, $\dmove$,
and a Boolean flag indicating whether the current best snapshot satisfies
the validity gate.

\paragraph{System prompt and advisory scaffold.}
The system prompt encodes domain knowledge in four parts: (i)~a parameter
description with physical bounds, (ii)~a four-stage advisory schedule
(Table~\ref{tab:schedule}) that acts as a pacing hint, (iii)~critical
timing rules for $\betaH$ advancement indexed to the budget-consumed
fraction, and (iv)~explicit guidance on when not to raise~$\betaH$.
The advisory schedule is \emph{not} enforced as a hard constraint; the
LLM may deviate at any iteration, and in practice, it does so when the observed grayness trajectory indicates the problem is not following the
nominal pacing.

\begin{table}[!htbp]
  \caption{Four-stage advisory parameter schedule embedded in the LLM
    system prompt. All four continuous parameters are advisory targets, not hard
    constraints; the agent outputs exact numeric values that may deviate
    based on the observed solver state. Budget fractions are expressed as a
    percentage of the total main-loop iteration count~$N$}
  \label{tab:schedule}
  \small
  \setlength{\tabcolsep}{4pt}
  \begin{tabular}{@{}lcccc@{}}
    \toprule
    \textbf{Stage} & \textbf{Budget} &
    $\boldsymbol{\penal}$ & $\boldsymbol{\betaH}$ &
    $\boldsymbol{\rmin}$ / $\boldsymbol{\dmove}$ \\
    \midrule
    1.\ Exploration  & $0$--$8\%$   & $1.0$--$2.0$ & $1.0$     & $1.50$ / $0.20$ \\
    2.\ Penalization & $8$--$50\%$  & $2.0$--$4.5$ & $1$--$4$  & $1.35$ / $0.15$ \\
    3.\ Sharpening   & $50$--$75\%$ & $4.5$        & $4$--$16$ & $1.25$ / $0.08$ \\
    4.\ Converge     & $75$--$100\%$& $4.5$        & $32$      & $1.20$ / $0.05$ \\
    \midrule
    Tail (fixed)     & $+40$ iters  & $4.5$        & $32$      & $1.20$ / $0.05$ \\
    \bottomrule
  \end{tabular}
\end{table}

\paragraph{Direct Numeric Control output.}
The LLM outputs a single JSON object at each call:
\begin{equation}
  a_t \;=\; \bigl\{\,
  \penal_t,\;\betaH_t,\;{r_{\min}}_t,\;\dmove_t,\;\mathrm{restart}_t,\;
  \mathrm{note}_t
  \,\bigr\},
  \label{eq:action}
\end{equation}
where $\mathrm{note}_t$ is a one-line rationale retained for auditing but
not fed back to the solver.
All numeric outputs are clamped to the ranges in Table~\ref{tab:params}
by a deterministic safety layer before application.
Additionally, $\rmin$ is enforced to be non-increasing across iterations
(monotone constraint), and the restart flag is ignored unless the best snapshot is valid.

\begin{table}[!htbp]
  \caption{LLM-controlled solver parameters, admissible ranges, and the
    hard grayness gate applied by the safety layer before each action is
    committed to the solver}
  \label{tab:params}
  \small
  \setlength{\tabcolsep}{4pt}
  \begin{tabular}{@{}llcl@{}}
    \toprule
    \textbf{Parameter} & \textbf{Symbol} & \textbf{Range} & \textbf{Role} \\
    \midrule
    Penalization exp. & $\penal$   & $[1.0,\;5.0]$   & SIMP void penalty \\
    Heaviside sharp.  & $\betaH$   & $[1.0,\;64.0]$  & Binarization \\
    Filter radius     & $\rmin$    & $[1.1,\;4.0]$   & Length scale ($\downarrow$) \\
    OC move limit     & $\dmove$   & $[0.03,\;0.40]$ & Density step \\
    Restart flag      & ---         & $\{\texttt{T},\texttt{F}\}$ & Reload snapshot \\
    \midrule
    \multicolumn{4}{@{}l}{\textit{Hard gate:} $\betaH \leq 8.0$ while
      $\Gray > 0.20$ (not overridable)} \\
    \bottomrule
  \end{tabular}
\end{table}

\paragraph{Grayness gate.}
The safety layer caps $\betaH \leq 8.0$ whenever $\Gray > 0.20$,
regardless of what the LLM requests.
This gate encodes an empirically established principle: aggressive Heaviside
sharpening while substantial intermediate-density material remains locks the
topology into a poor local minimum that the \OC{} method cannot
escape~\citep{Lazarov2016,Sigmund2007}.
Note that the system prompt gives the LLM a softer advisory
($\betaH < 16$ while $\Gray > 0.20$) to encourage patient behavior,
while the deterministic safety layer enforces the stricter hard cap
($\betaH \leq 8.0$); the cap holds even if the model ignores or
misinterprets the advisory.
The gate threshold ($0.20$) is itself a meta-optimizable constant, tuned by
the outer loop described in Sect.~\ref{sec:meta}.

\paragraph{Fallback schedule.}
When the Gemini API is unavailable or returns an invalid response, the agent falls back to a smooth deterministic ramp that follows the advisory
schedule in Table~\ref{tab:schedule}.
This fallback ensures the solver always advances; it is logged and excluded
from the primary results.

\subsection{Meta-Optimization: Outer Hyperparameter Loop}
\label{sec:meta}

\begin{figure}[!htbp]
  \centering
  \includegraphics[width=\linewidth]{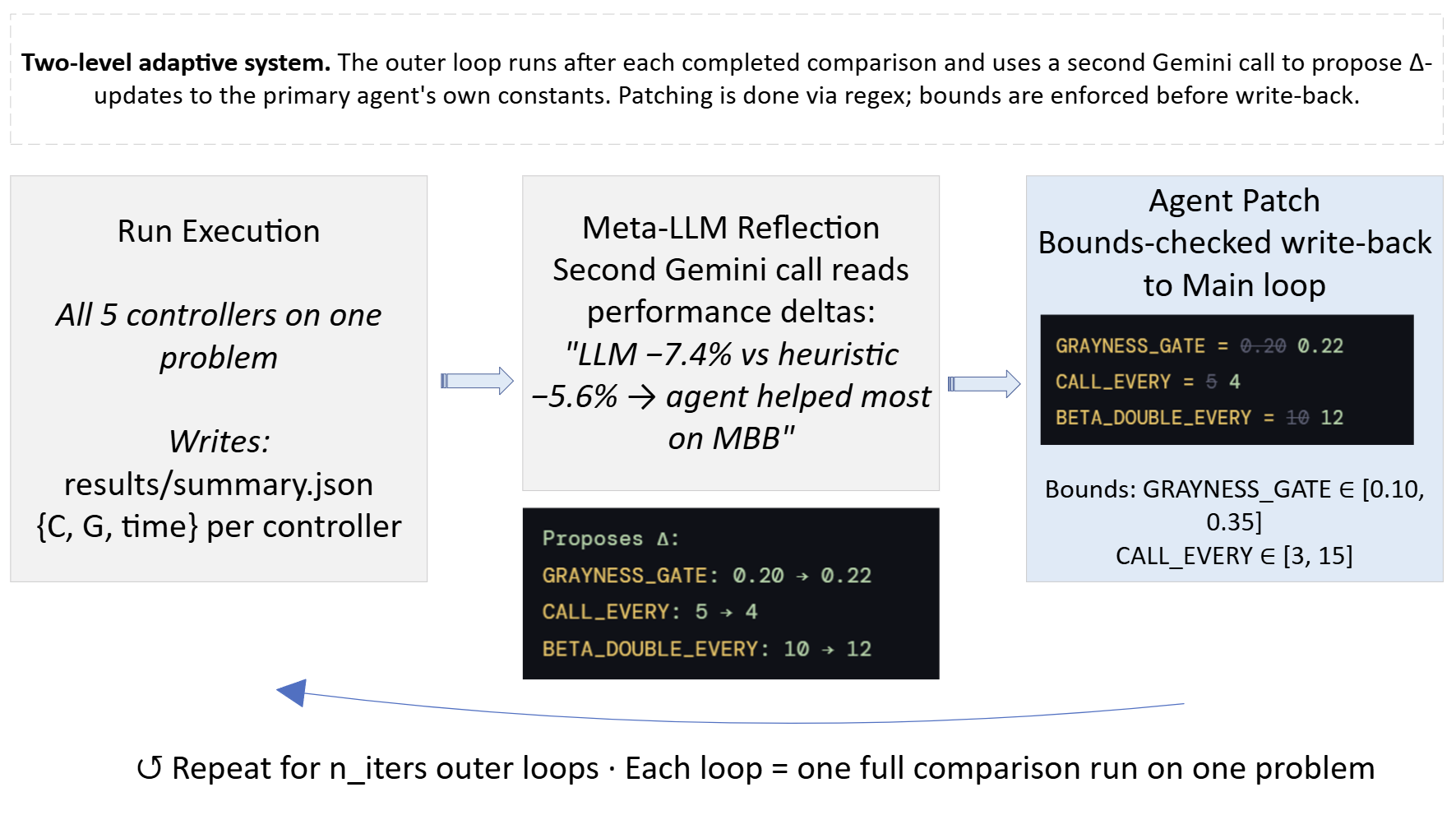}
  \caption{Meta-optimization outer loop. After each completed comparison
    run, the meta-optimizer reads the performance summary and submits a second Gemini call that reflects on the relative performance of all
    five controllers. The model proposes bounded $\Delta$-updates to the
    primary agent's own constants---the grayness-gate threshold, API call
    frequency, and fallback $\betaH$-doubling period
    (Table~\ref{tab:meta})---which are bounds-checked and written back to the agent configuration via automated patching. The outer loop repeats
    for $n_{\text{iters}}$ iterations, converging on agent hyperparameters
    that generalize across problem geometries}
  \label{fig:outerloop}
\end{figure}

The inner LLM agent adapts solver parameters within a single run.
A second level of adaptation---the meta-optimization loop---adjusts the
agent's own constants across successive runs (Fig.~\ref{fig:outerloop}).

Table~\ref{tab:meta} lists the six constants exposed to the outer loop with
their admissible bounds.
Hard bounds are enforced before write-back to prevent the meta-loop from
driving the agent into degenerate configurations.
To prevent overfitting to a single problem geometry, the outer loop is run
jointly over all three 2-D benchmarks ($n_{\text{iters}} = 5$ per problem),
cycling through cantilever, \MBB{} beam, and L-bracket in sequence.

\begin{table}[!htbp]
  \caption{Constants exposed to the meta-optimization outer loop, with
    hard bounds enforced before write-back. Default values are those
    obtained after convergence of the joint outer loop across all three
    2-D benchmarks; they are used for all primary experiments in
    Sect.~\ref{sec:results}}
  \label{tab:meta}
  \small
  \setlength{\tabcolsep}{4pt}
  \begin{tabular}{@{}llccc@{}}
    \toprule
    \textbf{Constant} & \textbf{Role} & \textbf{Lower} & \textbf{Upper} & \textbf{Default} \\
    \midrule
    \texttt{GRAYNESS\_GATE}  & $\betaH$-cap threshold   & $0.10$ & $0.35$ & $0.20$ \\
    \texttt{CALL\_EVERY}     & API call period           & $3$    & $15$   & $5$ \\
    \texttt{PENAL\_RAMP}     & Fallback ramp length      & $4$    & $20$   & $12$ \\
    \texttt{BETA\_DOUBLE}    & Fallback $\betaH$ period  & $5$    & $20$   & $10$ \\
    \texttt{PHASE.penal}     & Min iters in Stage~2      & $4$    & $25$   & $22$ \\
    \texttt{PHASE.sharp}     & Min iters in Stage~3      & $4$    & $25$   & $16$ \\
    \bottomrule
  \end{tabular}
\end{table}

\subsection{Baseline Controllers}
\label{sec:baselines}

Four controllers are compared against the LLM agent.

\paragraph{Fixed controller.}
Applies no continuation whatsoever; the solver runs with constant $\penal = 3$,
$\betaH = 1$, $\rmin = 1.5$, and $\dmove = 0.2$ for all $N$ iterations.
This is the true zero-intervention baseline; it reliably produces gray material
($\Gray \approx 0.10$--$0.15$) and higher compliance.

\paragraph{Three-field continuation.}
The standard academic schedule~\citep{Wang2011,Lazarov2016}: $\penal$
ramps linearly from $1.0$ to $4.5$ over $30$ iterations; $\betaH$ doubles
every $10$ iterations after the ramp; $\rmin$ is tightened in the late stage.
No state-dependent decisions are made.

\paragraph{Expert heuristic.}
A carefully hand-crafted schedule that mimics an experienced \SIMP{}
practitioner: step-wise $\penal$ ramp ($+0.75$ per $10$ iterations),
$\betaH$ raised only once $\penal \geq 3.0$, and a safe restart triggered
only when current compliance exceeds the best valid compliance by more
than $12\%$.
This controller represents the practical performance ceiling achievable
without run-time feedback.

\paragraph{Schedule-only (ablation).}
A fixed four-stage schedule based on the LLM agent's advisory scaffold
(Table~\ref{tab:schedule}), executed deterministically without any LLM
API calls.
Unlike the LLM agent, no solver state is observed, and no deviation from the pre-specified progression is possible.
This ablation isolates the contribution of the LLM's online adaptive
decisions from the contribution of the phase structure itself.

\subsection{Experimental Protocol and Standardized Sharpening Tail}
\label{sec:protocol}

\paragraph{Standardized tail.}
All four continuation controllers receive an \emph{identical} $40$-iteration
sharpening phase after the main loop concludes.
The tail restarts from the best valid intermediate snapshot
$\boldsymbol{\rho}^*$ (the density field achieving the lowest compliance while
satisfying the volume constraint) and applies $\penal = 4.5$, $\betaH = 32$,
$\rmin = 1.20$, $\dmove = 0.05$ for $40$ iterations.
Pilot runs confirmed that compliance changes negligibly beyond $40$ tail
iterations for every controller tested.
The fixed (no-continuation) controller does not receive a tail and serves
as the true zero-intervention reference.
Because the tail is a shared, fixed post-processing step applied identically
to the four continuation controllers, any compliance difference among them at run termination must originate exclusively from the quality of the
topology produced during the main exploration loop.

\paragraph{Benchmark problems.}
Experiments are conducted on three canonical 2-D benchmarks: the cantilever
beam (tip load, fixed left edge), the \MBB{} beam (three-point bending),
and the L-bracket (corner re-entrant geometry with stress concentration).
All use volume fraction $\vf = 0.40$, Poisson's ratio $\nu = 0.3$, and
Young's modulus ratio $\Emod_{\min}/\Emod_0 = 10^{-9}$.
Table~\ref{tab:benchmarks} summarises the mesh configurations.

\begin{table}[!htbp]
  \caption{Benchmark configurations. The primary results in
    Sect.~\ref{sec:results} use the $120{\times}60$ long preset.
    The fast preset is used during meta-optimization outer-loop iterations.
    3-D experiments use AMG-preconditioned conjugate-gradient via
    \texttt{pyamg}}
  \label{tab:benchmarks}
  \small
  \setlength{\tabcolsep}{4pt}
  \begin{tabular}{@{}llcccc@{}}
    \toprule
    \textbf{Problem} & \textbf{Preset} & \textbf{Mesh} &
    $\boldsymbol{N}$ & $\vf$ & \textbf{Primary?} \\
    \midrule
    \multirow{3}{*}{Cantilever}
      & Fast  & $60{\times}30$        & 100 & 0.40 & \\
      & Long  & $120{\times}60$       & 300 & 0.40 & $\checkmark$ \\
      & Hard  & $180{\times}90$       & 300 & 0.40 & \\
    \midrule
    \MBB{} beam      & Long  & $120{\times}60$          & 300 & 0.40 & $\checkmark$ \\
    L-bracket        & Long  & $120{\times}60$          & 300 & 0.40 & $\checkmark$ \\
    \midrule
    Cantilever (3-D) & ---   & $40{\times}20{\times}10$ & 300 & 0.40 & $\checkmark$ \\
    \MBB{} beam (3-D)& ---   & $40{\times}20{\times}10$ & 300 & 0.40 & $\checkmark$ \\
    \bottomrule
  \end{tabular}
\end{table}

\paragraph{Linear solver.}
For 2-D problems the stiffness system $\mathbf{K}\mathbf{U} = \mathbf{F}$
is solved by direct sparse factorization (SciPy).
For 3-D problems, an algebraic multigrid (\AMG{}) preconditioner
(\texttt{pyamg}) with conjugate-gradient iteration is
used~\citep{Bell2023PyAMG}.
The \AMG{} hierarchy is rebuilt only when the relative change in element
stiffness values exceeds $15\%$.

\paragraph{Evaluation metrics.}
The primary metric is the final compliance~$\Comp$ after the standardized
tail.
Secondary metrics are the final grayness~$\Gray$, wall-clock time, and the
iteration at which the best valid snapshot $\boldsymbol{\rho}^*$ is found.
Percentage improvements are quoted relative to the fixed baseline.
All results are reported as mean~$\pm$ standard deviation over
$n = 5$ independent random seeds.

\paragraph{Overall pipeline.}
The two-level system is illustrated in Fig.~\ref{fig:pipeline}.

\section{Results}
\label{sec:results}

\subsection{Primary 2-D Benchmarks}
\label{sec:results_2d}

Table~\ref{tab:results_main} summarises results on three benchmarks at
$120{\times}60$ resolution with $N=300$ main-loop iterations.
Final compliance is measured after the standardized $40$-iteration sharpening
tail; percentage changes are relative to the fixed baseline mean.

\begin{table}[!htbp]
  \caption{Primary 2-D results: mean$\,\pm\,$std final compliance~$\Comp$,
    grayness~$\Gray$, improvement versus the fixed baseline, wall-clock
    time, and the mean iteration at which the best valid snapshot entering the sharpening tail was found. \textbf{Bold}: best compliance per problem.
    All meshes: $120{\times}60$, $V_f=0.40$, $N=300$, $n=5$ seeds}
  \label{tab:results_main}
  \small\setlength{\tabcolsep}{4pt}
  \begin{tabular}{@{}llccccc@{}}
    \toprule
    \textbf{Problem} & \textbf{Controller} &
    $\bar{\Comp}\pm\sigma$ & $\bar{\Gray}$ &
    \textbf{vs.\ Fixed} & \textbf{Time (s)} & \textbf{Best iter} \\
    \midrule
    \multirow{5}{*}{Cantilever}
      & Fixed              & $80.978 \pm 0.000$ & $0.098$ & ---       & 384.9 & 300.0 \\
      & Three-field cont.\ & $76.786 \pm 0.000$ & $0.000$ & $-5.18\%$ & 337.4 &  83.0 \\
      & Expert heuristic   & $76.474 \pm 0.014$ & $0.000$ & $-5.56\%$ & 348.1 &  82.4 \\
      & Schedule only      & $81.330 \pm 0.000$ & $0.000$ & $+0.43\%$ & 289.0 & 267.6 \\
      & \textbf{LLM agent} & $\mathbf{76.407 \pm 0.021}$ & $\mathbf{0.000}$ & $\mathbf{-5.65\%}$ & 445.2 & 167.6 \\
    \midrule
    \multirow{5}{*}{MBB beam}
      & Fixed              & $102.281 \pm 0.000$ & $0.146$ & ---       & 425.3 & 300.0 \\
      & Three-field cont.\ & $96.759  \pm 0.010$ & $0.000$ & $-5.40\%$ & 396.4 &  75.2 \\
      & Expert heuristic   & $96.500  \pm 0.000$ & $0.000$ & $-5.65\%$ & 366.6 & 287.4 \\
      & Schedule only      & $103.393 \pm 0.001$ & $0.000$ & $+1.09\%$ & 344.8 &  67.0 \\
      & \textbf{LLM agent} & $\mathbf{94.646 \pm 0.050}$ & $\mathbf{0.000}$ & $\mathbf{-7.46\%}$ & 562.4 & 134.4 \\
    \midrule
    \multirow{5}{*}{L-bracket}
      & Fixed              & $46.325 \pm 0.000$ & $0.113$ & ---       & 343.3 & 300.0 \\
      & Three-field cont.\ & $43.799 \pm 0.007$ & $0.000$ & $-5.45\%$ & 268.5 &  81.8 \\
      & Expert heuristic   & $43.439 \pm 0.001$ & $0.000$ & $-6.23\%$ & 282.9 &  93.0 \\
      & Schedule only      & $44.872 \pm 0.000$ & $0.000$ & $-3.14\%$ & 248.7 & 262.4 \\
      & \textbf{LLM agent} & $\mathbf{43.028 \pm 0.000}$ & $\mathbf{0.000}$ & $\mathbf{-7.12\%}$ & 150.1 & 112.0 \\
    \bottomrule
  \end{tabular}
\end{table}

\begin{figure}[!htbp]
  \centering
  \includegraphics[width=\textwidth]{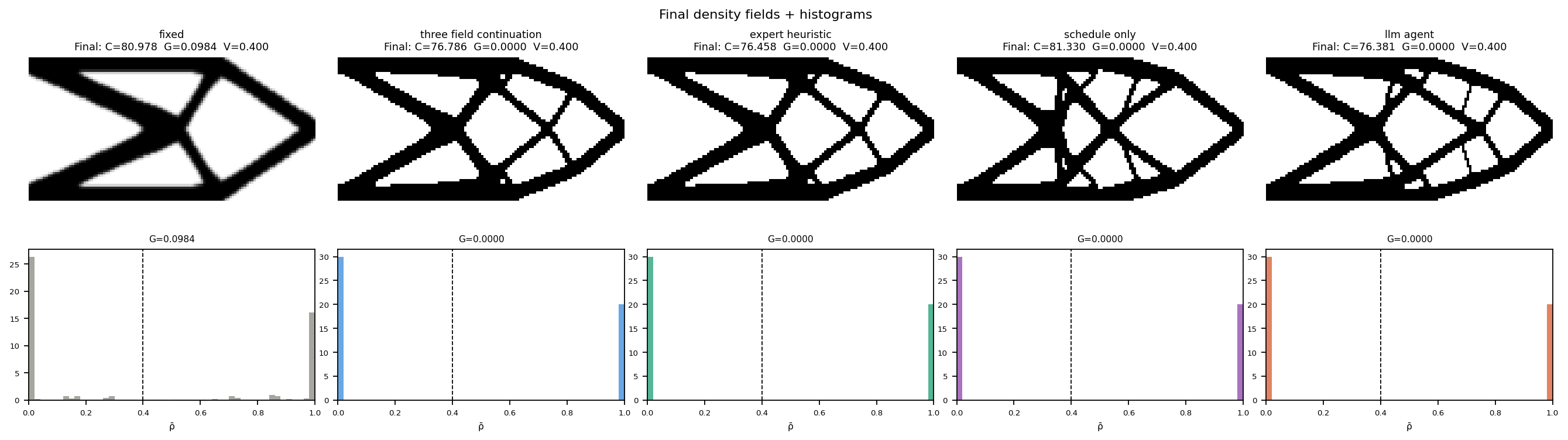}\\[6pt]
  \includegraphics[width=\textwidth]{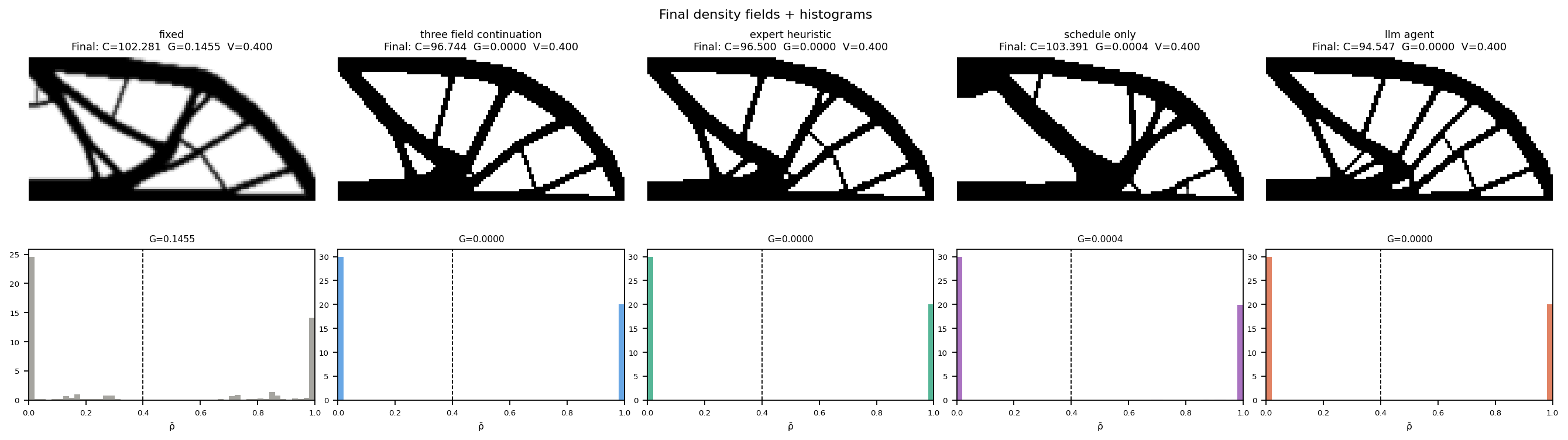}\\[6pt]
  \includegraphics[width=\textwidth]{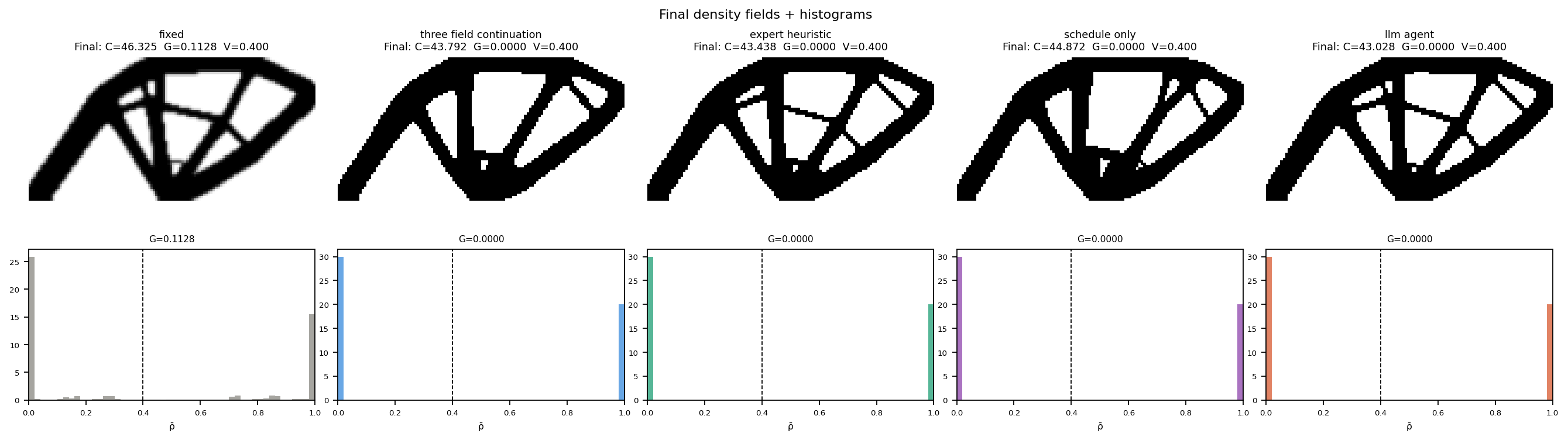}
  \caption{Final physical density fields~$\rhotil$ and element-density
    histograms for all five controllers ($120{\times}60$, $V_f=0.40$,
    representative run). \textit{Top row}: cantilever. \textit{Middle row}:
    \MBB{} beam. \textit{Bottom row}: L-bracket. The fixed controller retains
    substantial gray material ($\Gray>0$), visible as blurred member
    boundaries and a non-bimodal histogram. All continuation controllers
    achieve $\Gray=0.000$ with bimodal histograms concentrated at
    $\rhotil\in\{0,1\}$. Column order (left to right): fixed,
    three-field continuation, expert heuristic, schedule-only, LLM agent}
  \label{fig:designs_2d}
\end{figure}

\paragraph{Overall ranking.}
The LLM agent achieves the lowest mean compliance on all three benchmarks,
outperforming the next-best controller (expert heuristic) by
$0.067$ ($0.09\%$) on cantilever, $1.854$ ($1.92\%$) on the \MBB{} beam,
and $0.411$ ($0.95\%$) on the L-bracket.
All continuation controllers and the LLM agent converge to fully binary
topologies ($\Gray=0.000$), while the fixed baseline retains substantial
gray material ($\Gray\approx0.10$--$0.15$), confirming that continuation
is necessary for quality binarization.
Representative final density fields and element-density histograms for all
five controllers are shown in Fig.~\ref{fig:designs_2d}.

\begin{figure}[!htbp]
  \centering
  \includegraphics[width=\linewidth,height=0.28\textheight,keepaspectratio]{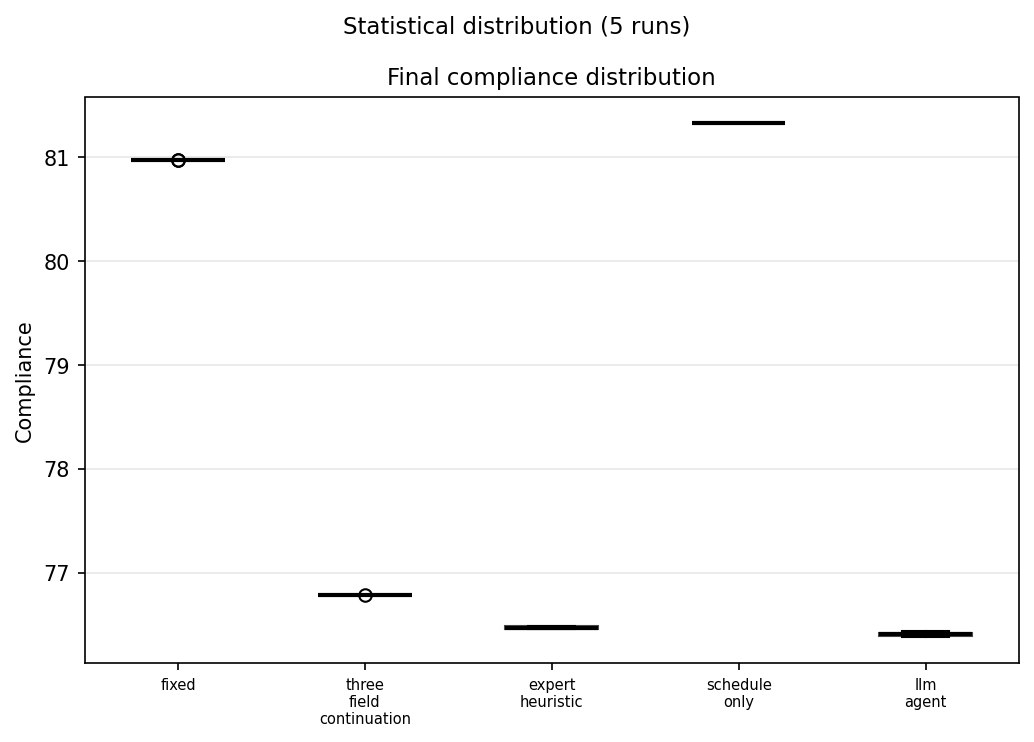}\\[2pt]
  \includegraphics[width=\linewidth,height=0.28\textheight,keepaspectratio]{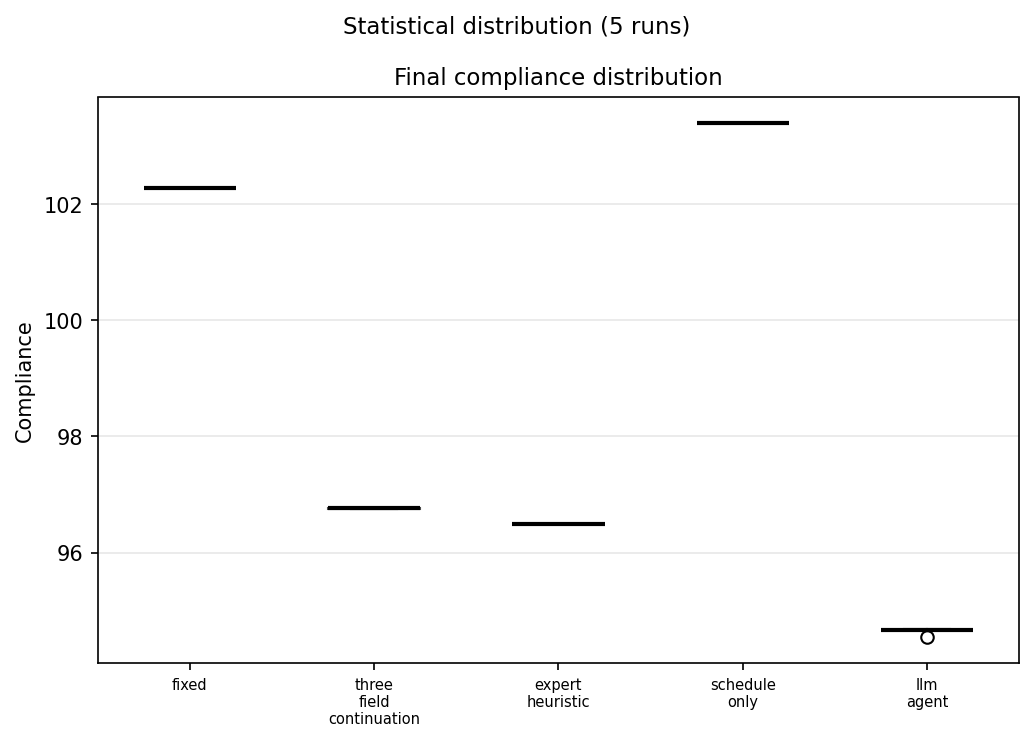}\\[2pt]
  \includegraphics[width=\linewidth,height=0.28\textheight,keepaspectratio]{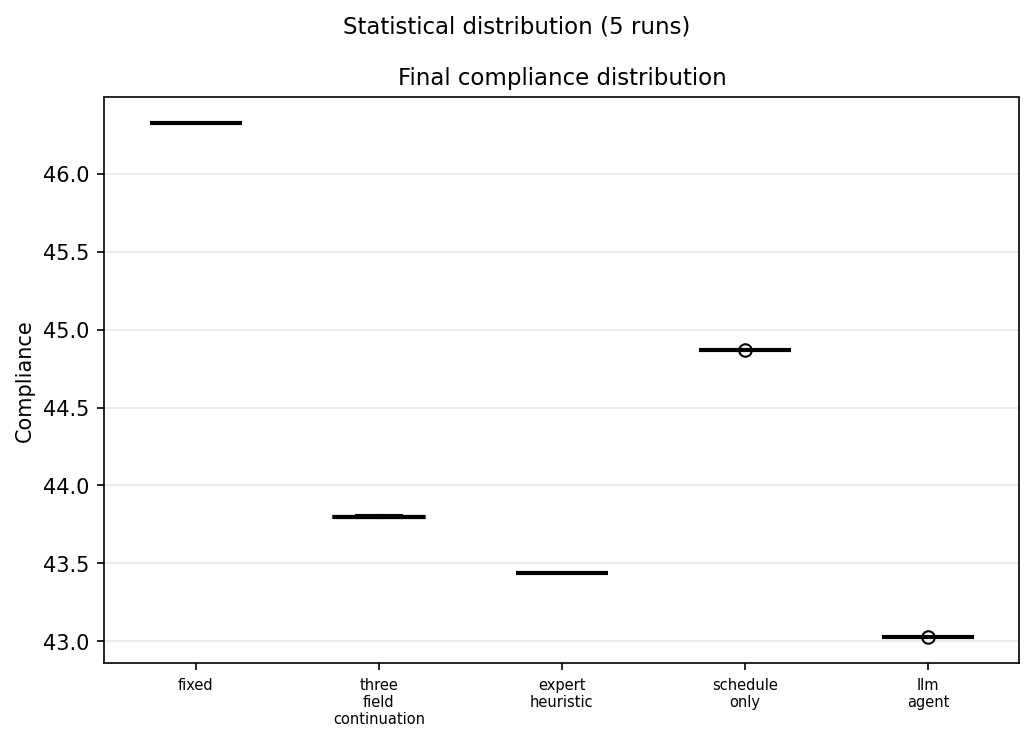}
  \caption{Final compliance distributions ($n=5$ runs) for all five
    controllers on cantilever (top), \MBB{} beam (middle), and L-bracket
    (bottom). Each panel plots the mean (filled circle) and $\pm3\sigma$
    interval. The LLM agent achieves the lowest mean on every problem with
    narrow spread. The schedule-only controller underperforms the fixed
    baseline on cantilever ($+0.43\%$) and \MBB{} beam ($+1.09\%$),
    confirming that rigid phase-structure adherence without state observation
    is actively harmful}
  \label{fig:statbox}
\end{figure}

\begin{figure}[!htbp]
  \centering
  \includegraphics[width=\linewidth]{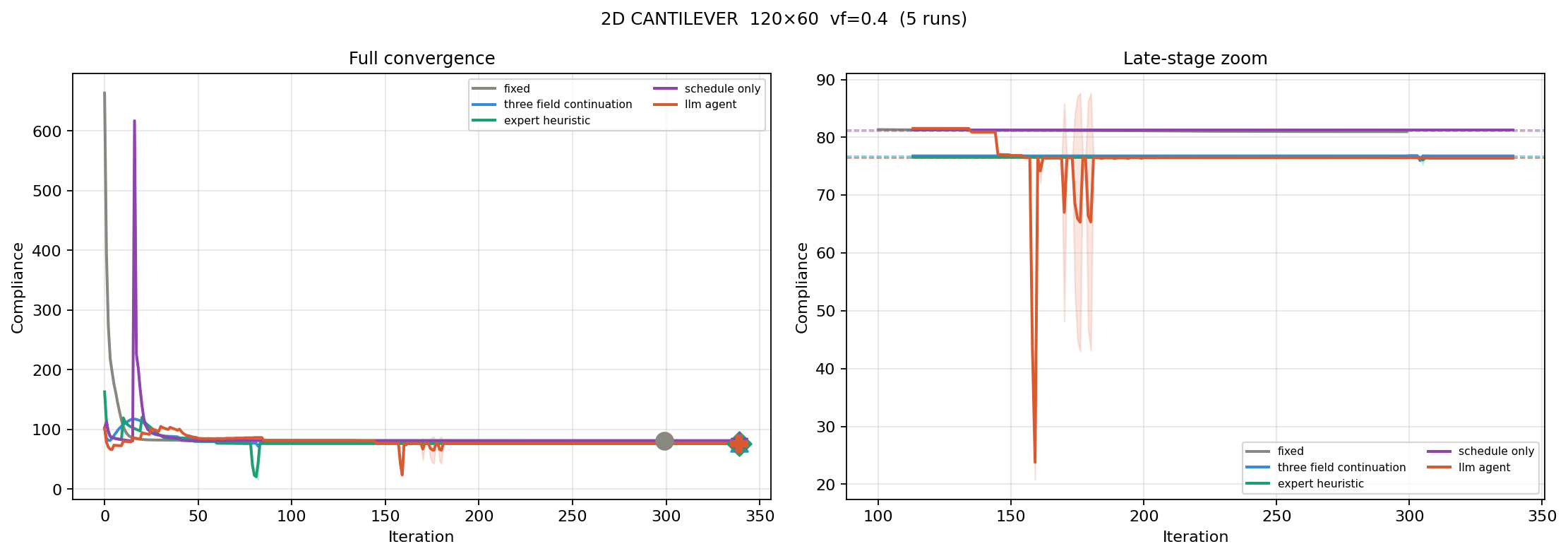}
  \caption{Compliance convergence curves for the cantilever benchmark
    ($120{\times}60$, $n=5$ seeds, shaded band = $\pm1\sigma$).
    \textit{Left}: full 340-iteration trace (300 main loop $+$ 40 tail).
    The large compliance spike near iteration~$160$ in the LLM agent trace
    (orange) corresponds to aggressive Heaviside sharpening; the subsequent restart recovers compliance well below the heuristic controllers.
    \textit{Right}: late-stage zoom (iterations~$100$--$350$), confirming
    that the LLM agent's final compliance is consistently lower than all
    alternatives}
  \label{fig:curves_canti}
\end{figure}

\begin{figure}[!htbp]
  \centering
  \includegraphics[width=\linewidth]{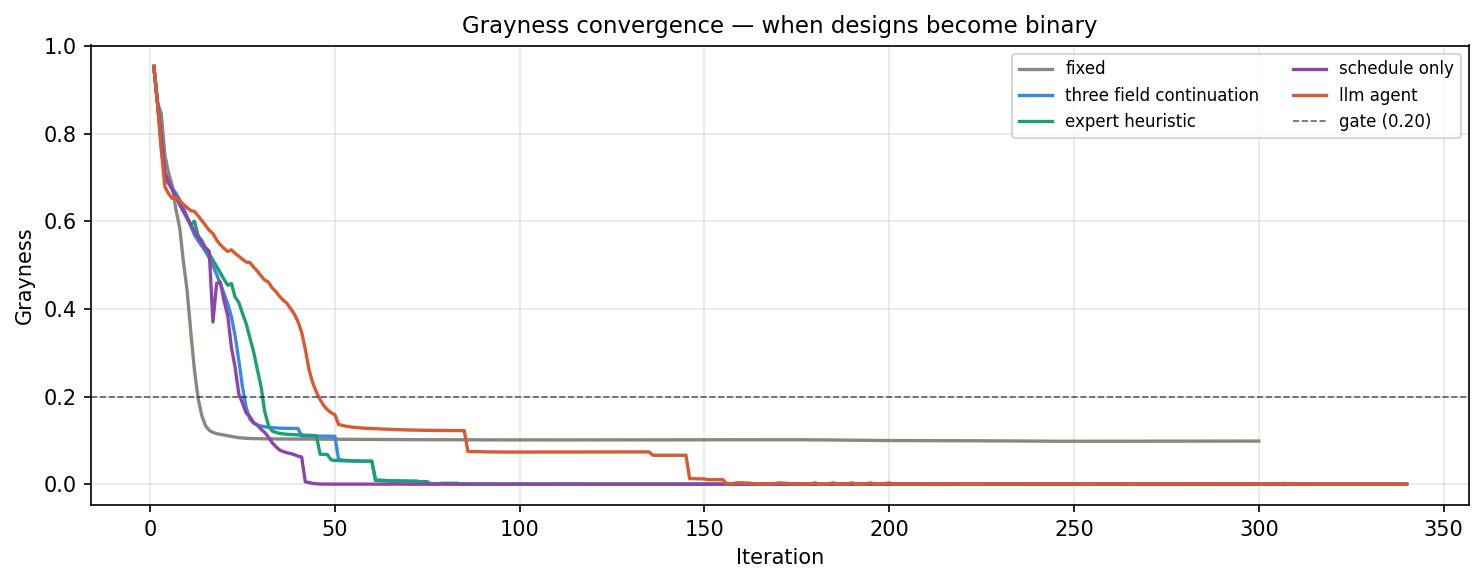}
  \caption{Grayness convergence $\Gray(t)$ for the cantilever benchmark
    ($120{\times}60$, representative run). The dashed horizontal line marks
    the grayness gate threshold ($\Gray=0.20$). The fixed controller
    plateaus at $\Gray\approx0.10$ and never achieves full binarization.
    The LLM agent crosses the gate threshold around iteration~$150$---substantially
    later than three-field continuation ($\sim50$) and the expert heuristic
    ($\sim60$)---confirming its more patient penalization strategy}
  \label{fig:grayness_canti}
\end{figure}

\paragraph{Statistical robustness.}
The LLM agent shows the smallest non-zero standard deviation
($\sigma=0.021$ on cantilever, $\sigma=0.050$ on \MBB{} beam),
indicating low run-to-run variance despite the stochastic nature of
LLM sampling.
(The fixed and schedule-only controllers are fully deterministic and
therefore produce $\sigma=0.000$ across seeds; three-field continuation
and the expert heuristic show near-zero variance due to minor
floating-point path differences.)
Figure~\ref{fig:statbox} shows the full compliance distribution across all
five runs.
The compliance convergence curves and parameter trajectories are presented
in Figs.~\ref{fig:curves_canti}--\ref{fig:params_canti}; the key mechanistic
observation---that the LLM agent defers binarization substantially longer
than fixed-schedule controllers---is directly visible in these traces.

\paragraph{Compliance and grayness convergence.}
Figure~\ref{fig:curves_canti} shows full compliance histories for the
cantilever benchmark.
The left panel reveals the sharp compliance spikes that the LLM agent
deliberately induces by triggering restarts to recover from local minima,
visible as the large transient near iteration~$160$.
The right (late-stage zoom) panel confirms that after each restart, the LLM agent recovers and drives compliance below all heuristic controllers.
Figure~\ref{fig:grayness_canti} shows the grayness convergence; the LLM agent
maintains $\Gray>0.20$ until approximately iteration~$150$, substantially
later than the fixed-schedule controllers, which cross the gate threshold
around iteration~$50$--$80$.
This directly illustrates the mechanistic claim: the LLM agent does not
prematurely binarize, allowing the topology to form freely under lower
$\betaH$ before committing to sharpening.

\subsection{LLM Agent Decision Trace}
\label{sec:llm_trace}

\begin{figure}[!htbp]
  \centering
  \includegraphics[width=\linewidth]{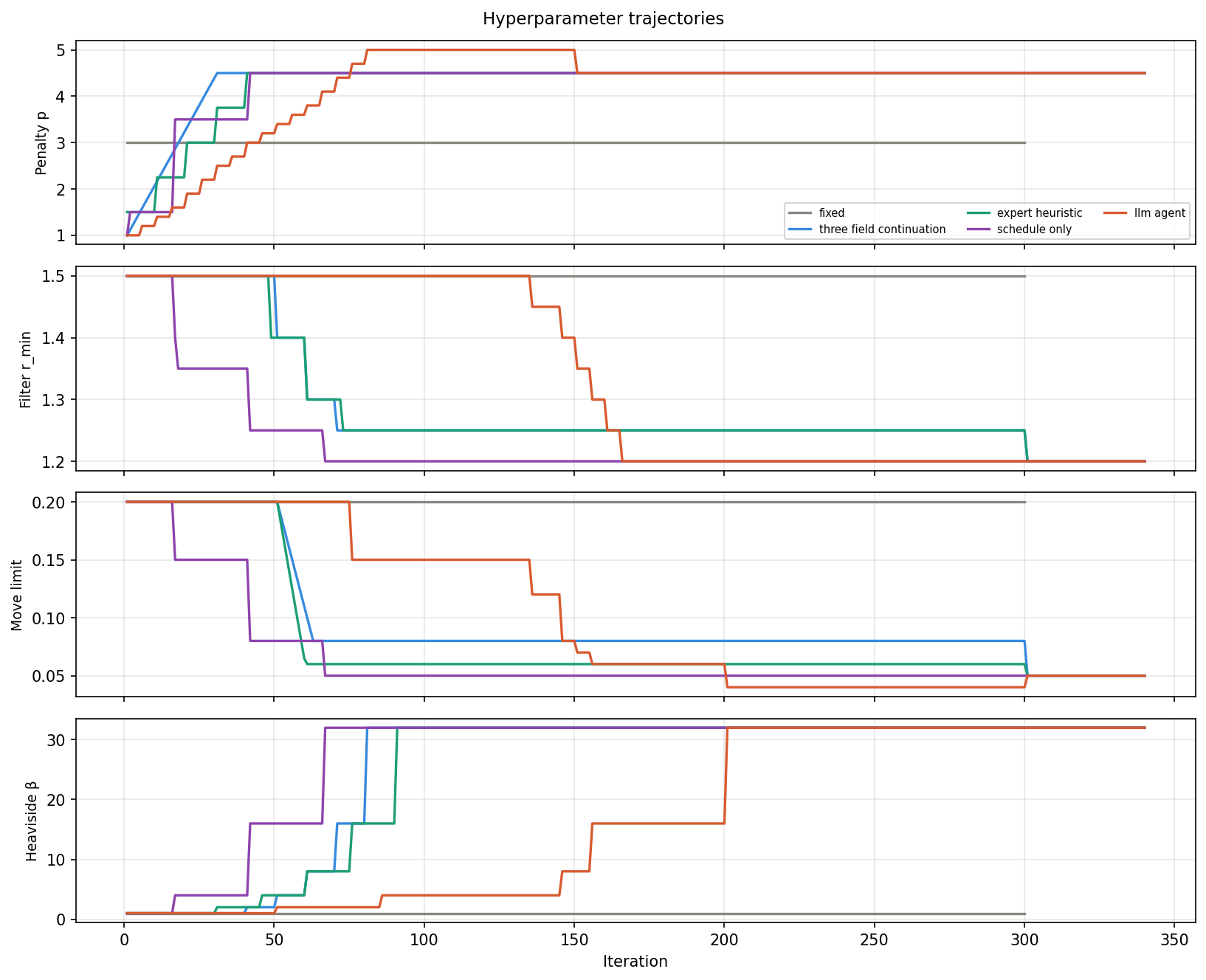}
  \caption{Hyperparameter trajectories for all five controllers on the
    cantilever benchmark ($120{\times}60$, representative run). Four
    subpanels show $\penal(t)$, $\rmin(t)$, $\dmove(t)$, and $\betaH(t)$
    over the full 340-iteration run (300 main $+$ 40 tail). The LLM agent
    (orange) maintains $\betaH=1$--$4$ for the first $\approx\!145$
    iterations before escalating to $\betaH=32$, compared to the
    schedule-only (purple) and three-field continuation (blue), which reach
    $\betaH\geq16$ by iteration~80}
  \label{fig:params_canti}
\end{figure}

To make the LLM agent's behavior interpretable, we present a detailed
trace of every API call during a representative cantilever run
(run~4 of~5, final compliance $\Comp=76.381$).
Table~\ref{tab:llm_trace} excerpts representative calls from each stage,
showing the parameter values output, the agent's one-line rationale, and
whether any safety constraint was active.
Figure~\ref{fig:params_canti} shows the full hyperparameter trajectories for
all five controllers.

{\small\setlength{\tabcolsep}{3pt}
\begin{longtable}{@{}rcccccp{5.2cm}@{}}
  \caption{Representative LLM agent calls during a cantilever run
    (run~4, $N=300$, final $\Comp=76.381$).
    restart~$=\checkmark$ indicates a snapshot reload was requested
    \emph{and} executed. The \textit{Gate} column marks iterations where
    $\Gray > 0.20$, indicating that the grayness gate was active}
  \label{tab:llm_trace} \\
  \toprule
  \textbf{Iter.} & $\penal$ & $\betaH$ & $\rmin$ & $\dmove$ &
  \textbf{Gate} & \textbf{LLM rationale (verbatim)} \\
  \midrule
  \endfirsthead
  \multicolumn{7}{@{}l}{\tablename\ \thetable{} -- \textit{continued}} \\[2pt]
  \toprule
  \textbf{Iter.} & $\penal$ & $\betaH$ & $\rmin$ & $\dmove$ &
  \textbf{Gate} & \textbf{LLM rationale (verbatim)} \\
  \midrule
  \endhead
  \midrule
  \multicolumn{7}{r@{}}{\textit{Continued on next page}} \\
  \endfoot
  \bottomrule
  \\[-6pt]
  \multicolumn{7}{@{}p{\linewidth}@{}}{%
    \footnotesize $^\dagger$At iterations 90--130, $\Gray$ remained
    marginally above the gate threshold ($0.20 < \Gray \lesssim 0.24$);
    the deterministic safety layer capped $\betaH \leq 8$ accordingly.
    The model's characterization of grayness as ``low'' reflects its own
    interpretation of the observation vector rather than a comparison
    against the gate threshold, which the model does not observe directly.
  } \\
  \endlastfoot
  \multicolumn{7}{@{}l}{\textit{Stage 1 - Exploration (budget 0--8\%)}} \\[1pt]
  5  & 1.2 & 1.0 & 1.50 & 0.20 & --- & ``Stage 1 exploration, slowly increasing penal.'' \\
  15 & 1.6 & 1.0 & 1.50 & 0.20 & --- & ``Increasing penal slightly to encourage feature definition.'' \\
  20 & 1.9 & 1.0 & 1.50 & 0.20 & --- & ``Increasing penal to encourage structural definition.'' \\[3pt]
  \multicolumn{7}{@{}l}{\textit{Stage 2 - Penalization (budget 8--50\%): patient $p$ ramp, $\betaH$ held low}} \\[1pt]
  25 & 2.2 & 1.0 & 1.50 & 0.20 & --- & ``Entering Stage 2: increasing penalty to drive grayness reduction.'' \\
  40 & 3.0 & 1.0 & 1.50 & 0.20 & --- & ``Increasing penal to address stagnation and high grayness.'' \\
  50 & 3.4 & 2.0 & 1.50 & 0.20 & --- & ``Increasing penal and beta to drive binarization.'' \\
  80 & 5.0 & 2.0 & 1.50 & 0.15 & --- & ``Grayness is low; increasing penal to 5.0.'' \\
  85 & 5.0 & 4.0 & 1.50 & 0.15 & --- & ``Grayness is low; increasing $\beta$ to 4.0.'' \\[3pt]
  \multicolumn{7}{@{}l}{\textit{Stage 2 hold (budget 28--45\%): grayness gate active ($\Gray>0.20$), $\betaH$ held at 4}} \\[1pt]
  90  & 5.0 & 4.0 & 1.50 & 0.15 & $\checkmark$ & ``Grayness is low; maintaining Stage 2 parameters.''$^\dagger$ \\
  120 & 5.0 & 4.0 & 1.50 & 0.15 & $\checkmark$ & ``Maintaining $p=5.0$, $\beta=4.0$ to stabilize.''$^\dagger$ \\
  130 & 5.0 & 4.0 & 1.50 & 0.15 & $\checkmark$ & ``Maintaining current parameters to ensure stability.''$^\dagger$ \\[3pt]
  \multicolumn{7}{@{}l}{\textit{Stage 3 - Sharpening (budget 45--55\%): $\betaH$ ramp begins}} \\[1pt]
  135 & 5.0 & 4.0  & 1.45 & 0.12 & --- & ``Entering Stage 3: reducing $r_{\min}$ and move.'' \\
  145 & 5.0 & 8.0  & 1.40 & 0.08 & --- & ``Increasing $\beta$ to 8.0, reducing $r_{\min}$/move.'' \\
  155 & 4.5 & 16.0 & 1.30 & 0.06 & --- & ``Entering sharpening: increasing $\beta$ and reducing $r_{\min}$.'' \\
  165 & 4.5 & 16.0 & 1.20 & 0.06 & --- & ``$r_{\min}$ reduced to 1.20; compliance spike detected --- restart.'' \\[3pt]
  \multicolumn{7}{@{}l}{\textit{Stage 4 - Converge (budget 67--100\%): $\betaH=32$, repeated restarts}} \\[1pt]
  200 & 4.5 & 32.0 & 1.20 & 0.04 & --- & ``Compliance spike detected; restarting to recover from local minimum.'' \\
  225 & 4.5 & 32.0 & 1.20 & 0.04 & --- & ``Compliance spike detected; restarting to recover best valid state.'' \\
  300 & 4.5 & 32.0 & 1.20 & 0.04 & --- & ``Compliance spike extreme; restarting to recover best valid state.'' \\
\end{longtable}
}

Table~\ref{tab:llm_trace} reveals four distinct behavioral phases that the agent navigates autonomously, without being told which phase it is in.

\textit{Stage~1 (iterations 1--24).}
The agent issues slow, conservative $\penal$ increments ($1.2\to1.9$) while
holding $\betaH=1$ and $\rmin=1.50$, allowing the optimizer to form topology
freely with minimal binarization pressure.

\textit{Stage~2 (iterations 25--134): patient penalization, $\betaH$ gated.}
The agent ramps $\penal$ steadily from $2.2$ to $5.0$ while keeping
$\betaH\leq4.0$.
The grayness gate is active for approximately $45$ iterations
(iterations 85--134), during which the model requests $\betaH$ advances
that are silently capped.
This long plateau---where the agent ``waits'' for the topology to consolidate
under high penalization before permitting Heaviside sharpening---is the
primary behavioral difference from fixed-schedule controllers.
Three-field continuation and the expert heuristic cross their equivalent
binarization threshold around iterations 50--60; the LLM agent defers
this by $\approx\!70$--$90$ iterations.

\textit{Stage~3 (iterations 135--199): controlled $\betaH$ escalation.}
Once grayness drops sufficiently, the agent escalates $\betaH$ from
$4\to8\to16$ over $\approx\!20$ iterations, simultaneously reducing $\rmin$
($1.50\to1.20$) and $\dmove$ ($0.15\to0.06$).
The compliance spike visible near iteration~$160$ in Fig.~\ref{fig:curves_canti}
is a direct consequence of the $\betaH=16$ transition; the agent correctly
identifies this as a local minimum escape event and immediately requests a
snapshot restart.

\textit{Stage~4 (iterations 200--300): final convergence with persistent
restarts.}
With $\betaH=32$ and $\rmin=1.20$ locked in, the agent's primary function
becomes monitoring for compliance spikes and issuing restarts to reload the
best valid snapshot, actively preventing compliance drift.

\subsection{Ablation: Isolating the LLM Contribution}
\label{sec:ablation}

The schedule-only controller runs the identical four-stage phase structure
as the LLM agent---the same budget breakpoints and nominal parameter
targets---but without any LLM API calls.
If the agent's advantage were attributable entirely to the phase structure,
schedule-only should match or approach the LLM agent's compliance.
The data show the opposite.

On cantilever, schedule-only achieves $\bar{\Comp}=81.330$, which is
$+0.43\%$ \emph{above} the fixed baseline (worse than doing nothing) and
$+6.45\%$ above the LLM agent.
On the \MBB{} beam, schedule-only achieves $\bar{\Comp}=103.393$,
$+1.09\%$ above fixed, and $+9.24\%$ above the LLM agent.
On the L-bracket, schedule-only improves to $-3.14\%$ relative to fixed
but remains $+4.29\%$ above the LLM agent and $+3.3\%$ above the expert
heuristic.

Inspecting the parameter trajectories (Fig.~\ref{fig:params_canti})
reveals why schedule-only underperforms.
The schedule commits to $\betaH=32$ by iteration~80---the same point where the LLM agent is still holding $\betaH=2$.
This premature binarization crystallizes a sub-optimal
topology~\citep{StoMeSvanberg2001}: the best valid snapshot enters the tail
at iteration~$67.0$ on \MBB{} beam and $267.6$ on cantilever.
These results confirm: (i)~the four-stage advisory schedule alone provides
no consistent benefit over a fixed no-continuation controller; and
(ii)~the LLM's online state-conditioned decisions are the direct cause of
compliance improvements.

\subsection{3-D Benchmarks}
\label{sec:results_3d}

Table~\ref{tab:results_3d} shows results on two 3-D benchmarks at
$40{\times}20{\times}10$ resolution ($N=300$, $V_f=0.40$, $n=5$ seeds).

\begin{table}[!htbp]
  \caption{3-D results ($40{\times}20{\times}10$, $V_f=0.40$, $N=300$,
    $n=5$ seeds). \textbf{Bold}: best compliance per problem. The LLM agent
    achieves the lowest compliance on both benchmarks. Total improvement
    from continuation is $\approx\!15$--$18\%$ in 3-D versus $5$--$7\%$
    in 2-D, confirming that problems without an established optimal schedule
    benefit most from adaptive control}
  \label{tab:results_3d}
  \small\setlength{\tabcolsep}{3pt}
  \begin{tabular}{@{}llccccc@{}}
    \toprule
    \textbf{Problem} & \textbf{Controller}
      & $\bar{\Comp} \pm \sigma$ & $\bar{\Gray}$
      & \textbf{vs.\ Fixed} & \textbf{Time (s)} & \textbf{Best iter} \\
    \midrule
    \multirow{5}{*}{\rotatebox{90}{Cantilever}}
      & Fixed              & $4.860 \pm 0.000$ & $0.213$ & ---        & 1044 & 255 \\
      & Three-field cont.\ & $3.998 \pm 0.001$ & $0.000$ & $-17.74\%$ &  994 &  75 \\
      & Expert heuristic   & $3.999 \pm 0.001$ & $0.000$ & $-17.73\%$ &  933 &  94 \\
      & Schedule only      & $4.030 \pm 0.000$ & $0.000$ & $-17.09\%$ &  746 &  68 \\
      & \textbf{LLM agent} & $\mathbf{3.981 \pm 0.000}$ & $\mathbf{0.000}$ & $\mathbf{-18.09\%}$ & 1198 & 157 \\
    \midrule
    \multirow{5}{*}{\rotatebox{90}{\MBB{} beam}}
      & Fixed              & $6.429 \pm 0.000$ & $0.260$ & ---        & 1014 & ---\textsuperscript{$\ast$} \\
      & Three-field cont.\ & $5.480 \pm 0.001$ & $0.000$ & $-14.77\%$ & 1059 &  75 \\
      & Expert heuristic   & $5.442 \pm 0.001$ & $0.000$ & $-15.35\%$ & 1173 &  81 \\
      & Schedule only      & $5.517 \pm 0.000$ & $0.000$ & $-14.18\%$ & 1136 & 167 \\
      & \textbf{LLM agent} & $\mathbf{5.430 \pm 0.001}$ & $\mathbf{0.000}$ & $\mathbf{-15.55\%}$ & 1685 & 149 \\
    \bottomrule
    \\[-6pt]
    \multicolumn{7}{@{}p{0.95\linewidth}@{}}{%
      \footnotesize $^\ast$The fixed controller with $\penal=3$ and
      $\betaH=1$ never satisfies the validity gate ($\penal\geq3.0$ and
      $\Gray<0.25$) on this problem, so no valid best snapshot is recorded.
    }
  \end{tabular}
\end{table}

\begin{figure}[!htbp]
  \centering
  \includegraphics[width=\linewidth]{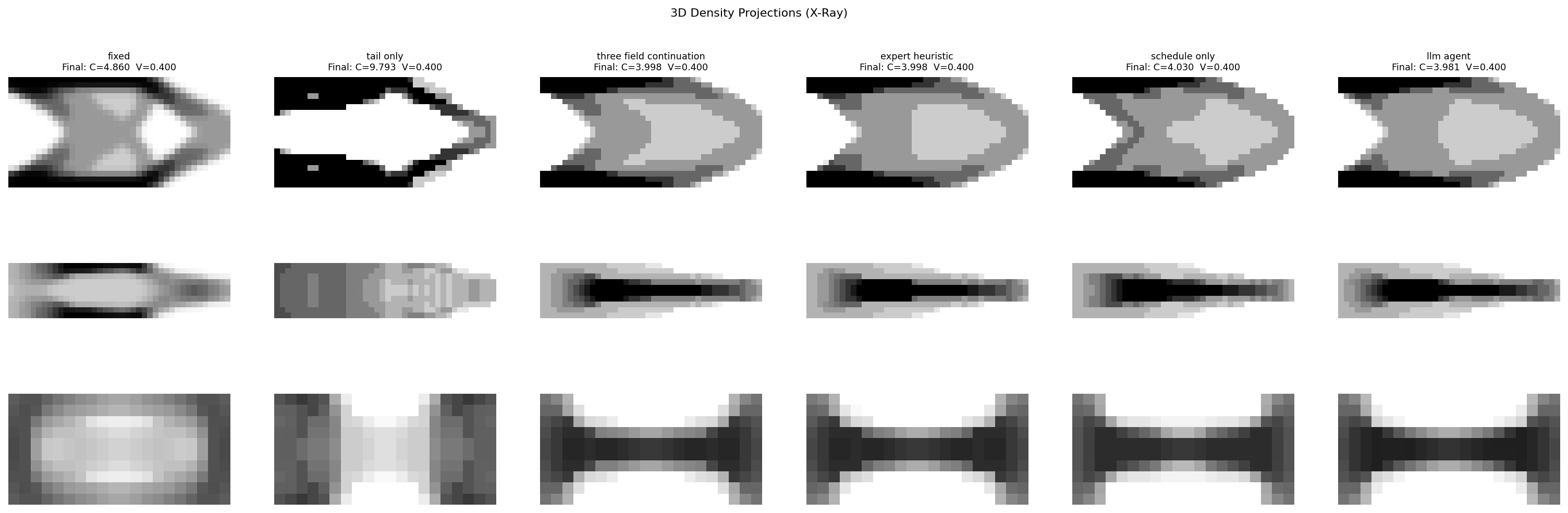}
  \caption{3-D density projections (X-ray view) for the $40{\times}20{\times}10$
    cantilever ($n=5$ seeds, representative run). Columns: fixed, tail-only,
    three-field continuation, expert heuristic, schedule-only, LLM agent.
    Rows: $xy$, $xz$, $yz$ projection directions. The fixed controller shows
    pervasive diffuse shading ($\Gray=0.213$); all continuation controllers
    produce binary structures}
  \label{fig:designs_3d_canti}
\end{figure}

\begin{figure}[!htbp]
  \centering
  \includegraphics[width=\linewidth]{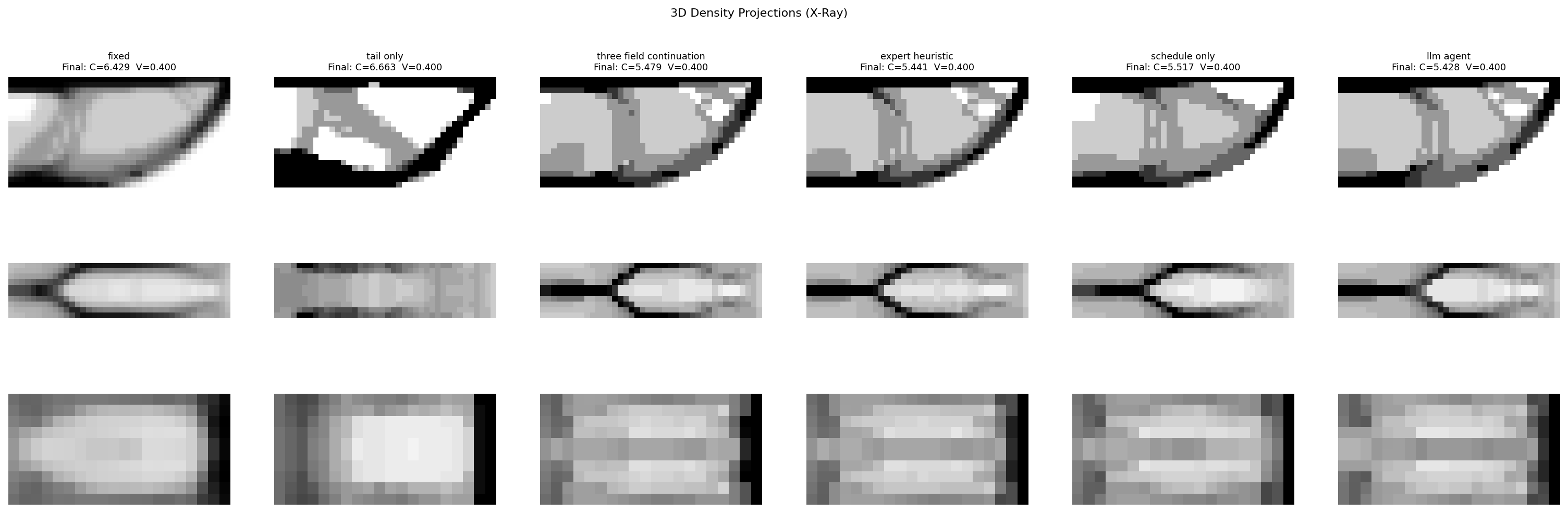}
  \caption{3-D density projections (X-ray view) for the $40{\times}20{\times}10$
    \MBB{} beam ($n=5$ seeds, representative run). The LLM agent (rightmost,
    $\Comp=5.428$) shows a more compact load-carrying core than the heuristic
    controllers}
  \label{fig:designs_3d_mbb}
\end{figure}

The LLM agent achieves $\bar{\Comp}=3.981$ ($-18.09\%$ vs.\ fixed) on the
3-D cantilever and $\bar{\Comp}=5.430$ ($-15.55\%$ vs.\ fixed) on the 3-D
\MBB{} beam.
The total improvement from any continuation over the fixed controller is
$\approx\!18\%$ in 3-D versus only $5$--$7\%$ in 2-D, consistent with the
expectation that problems without an established optimal schedule benefit most from adaptive control.
The LLM agent's best valid snapshot is found at iteration~$149$ on the
\MBB{} beam---roughly twice as late as the three-field continuation
(iteration~$75$)---again demonstrating the patient penalization mechanism.
Figures~\ref{fig:designs_3d_canti} and~\ref{fig:designs_3d_mbb} show the
3-D density projections for all controllers on each problem.

\subsection{Tail-Only Ablation: Necessity of the Exploration Phase}
\label{sec:tail_ablation}

To confirm that the standardized sharpening tail cannot substitute for
main-loop exploration, we introduce a tail-only controller that performs
no exploration: the main loop runs with $\penal=1.0$ and $\betaH=1.0$
(below the validity gate $\penal\geq3.0$), so that no valid best snapshot
is recorded, and the tail starts from uniform density.

Table~\ref{tab:tail_ablation} reports results on the three 2-D benchmarks
at $60{\times}30$ resolution ($N=100$, $n=5$ seeds).
The tail-only controller achieves compliance $115$--$141\%$ \emph{worse}
than the fixed no-continuation baseline on all three problems, confirming
that $40$ iterations of aggressive sharpening from a blank slate are
entirely insufficient to form a competitive topology.
This establishes that the exploration phase is the primary determinant of
final solution quality.
Note that at the reduced $60{\times}30$ resolution with $N=100$ iterations, the LLM agent wins on cantilever but is narrowly outperformed on the \MBB{}
beam and L-bracket, consistent with the primary $120{\times}60$ results
(Table~\ref{tab:results_main}): the patient penalization strategy requires
sufficient iteration budget ($N=300$) to defer binarization meaningfully.

\begin{table}[!htbp]
  \caption{Tail-only ablation ($60{\times}30$, $V_f=0.40$, $N=100$, $n=5$
    seeds). The tail-only controller starts from uniform density (no
    exploration). Its extreme underperformance confirms that the main-loop exploration is essential. \textbf{Bold}: best compliance per problem}
  \label{tab:tail_ablation}
  \small\setlength{\tabcolsep}{3pt}
  \begin{tabular}{@{}llccc@{}}
    \toprule
    \textbf{Problem} & \textbf{Controller}
      & $\bar{\Comp} \pm \sigma$ & $\bar{\Gray}$
      & \textbf{vs.\ Fixed} \\
    \midrule
    \multirow{6}{*}{\rotatebox{90}{Cantilever}}
      & Fixed              & $88.047  \pm 0.000$  & $0.190$ & --- \\
      & Tail only          & $211.777 \pm 0.000$  & $0.000$ & $+140.5\%$ \\
      & Three-field cont.\ & $79.449  \pm 0.000$  & $0.000$ & $-9.76\%$ \\
      & Expert heuristic   & $79.486  \pm 0.000$  & $0.000$ & $-9.72\%$ \\
      & Schedule only      & $83.082  \pm 0.000$  & $0.000$ & $-5.64\%$ \\
      & \textbf{LLM agent} & $\mathbf{79.301 \pm 0.086}$ & $\mathbf{0.000}$ & $\mathbf{-9.93\%}$ \\
    \midrule
    \multirow{6}{*}{\rotatebox{90}{MBB beam}}
      & Fixed              & $109.723 \pm 0.000$  & $0.203$ & --- \\
      & Tail only          & $246.560 \pm 0.000$  & $0.000$ & $+124.7\%$ \\
      & \textbf{Three-field cont.}\ & $\mathbf{96.323 \pm 0.000}$ & $\mathbf{0.000}$ & $\mathbf{-12.21\%}$ \\
      & Expert heuristic   & $96.695  \pm 0.000$  & $0.000$ & $-11.87\%$ \\
      & Schedule only      & $99.638  \pm 0.000$  & $0.000$ & $-9.19\%$ \\
      & LLM agent          & $96.980  \pm 0.000$  & $0.000$ & $-11.61\%$ \\
    \midrule
    \multirow{6}{*}{\rotatebox{90}{L-bracket}}
      & Fixed              & $49.221  \pm 0.000$  & $0.186$ & --- \\
      & Tail only          & $105.645 \pm 0.000$  & $0.000$ & $+114.6\%$ \\
      & Three-field cont.\ & $43.518  \pm 0.000$  & $0.000$ & $-11.59\%$ \\
      & \textbf{Expert heuristic}   & $\mathbf{43.437 \pm 0.000}$ & $\mathbf{0.000}$ & $\mathbf{-11.75\%}$ \\
      & Schedule only      & $44.413  \pm 0.000$  & $0.000$ & $-9.77\%$ \\
      & LLM agent          & $43.537  \pm 0.000$  & $0.000$ & $-11.55\%$ \\
    \bottomrule
  \end{tabular}
\end{table}

\section{Discussion}
\label{sec:discussion}

\subsection{Why the LLM Outperforms Fixed Schedules: The Patient Penalization Mechanism}

The central empirical finding of this work is that the LLM agent
consistently locates its best valid intermediate snapshot later in the run
than any fixed-schedule controller---at iteration~$167.6$ on cantilever
versus $82.4$ for the expert heuristic, and at iteration~$134.4$ on the
\MBB{} beam versus $75.2$ for three-field continuation.
This timing difference is not incidental: it is the direct consequence of the grayness gate preventing premature $\betaH$ escalation.
By holding $\betaH\leq8$ while $\Gray>0.20$ (active for roughly 45--70
iterations in each run), the agent allows the OC optimizer to continue
reshaping the density field under moderate penalization rather than
crystallizing topology prematurely into a near-binary
form~\citep{Lazarov2016,Sigmund2007}.
The superior intermediate topology found during this extended penalization
phase---lower in compliance even before the sharpening tail---then becomes
the starting point for the standardized tail, which sharpens whichever
topology it receives.
This behavior is consistent with the curriculum learning insight that
premature difficulty increases harm convergence~\citep{BengioCurriculum2009}:
by deferring the hardest constraint (full binarization) until the optimizer
is ready, the agent avoids the local minima that fixed-schedule controllers
fall into.

Because the tail is identical for all controllers, the compliance ordering at run end directly reflects the quality of the exploration-phase topology: the agent's patient approach finds a better gray-topology basin that the tail can sharpen to a lower final compliance.
This mechanism also explains why the \emph{schedule-only} ablation fails.
The schedule-only controller applies the same budget-breakpoint logic but
without observing actual $\Gray(t)$, so it advances $\betaH$ at a fixed
iteration count regardless of whether the topology is ready.
On the \MBB{} beam, schedule-only's best snapshot enters the tail at
iteration~$67.0$---a premature commitment made before the topology has
consolidated---while the LLM agent waits until iteration~$134.4$.
The schedule-only final compliance of $103.393$ ($+1.09\%$ vs.\ the fixed
no-intervention baseline) confirms that a pre-specified schedule is not merely
ineffective but actively harmful when applied rigidly: the phase transitions lock in parameter progressions that do not match the actual grayness
trajectory of the run.

\subsection{Interpretability and Auditability}

A distinctive property of the \DNC{} interface is that every parameter decision is accompanied by a one-line rationale emitted by the model (see
Table~\ref{tab:llm_trace}).
Unlike neural surrogate controllers or reinforcement-learned
policies~\citep{Shahriari2016,Hutter2019,EimerDACBench2021}, the LLM agent
provides a natural-language account of each action: the rationale strings in
Table~\ref{tab:llm_trace} allow an engineer to audit the agent's logic at any
point in the run and to detect failure modes such as premature binarization
or miscalibrated restart triggers without inspecting the underlying model
weights.
This interpretability is particularly valuable in structural design contexts
where decisions must be justifiable, and where unexpected topologies can carry
safety implications.

The call trace also reveals a limitation: during Stage~4 (iterations
$200$--$300$), the agent repeatedly emits ``compliance spike detected;
restarting'' regardless of whether the actual compliance has spiked or
whether a restart will help.
Inspection of the compliance curves (Fig.~\ref{fig:curves_canti}) shows that many of these restarts are genuine recoveries from OC perturbation, but some
are redundant---the topology at that iteration was already binary and stable.
This suggests that the system prompt's spike-detection language is over-fitted
to early-stage behavior and could be refined to reduce unnecessary API calls in the converge stage.

A second interpretability concern is visible in Table~\ref{tab:llm_trace} during
the Stage~2 hold (iterations 90--130): the agent characterizes grayness as
``low'' even though $\Gray$ remains above the gate threshold
($0.20 < \Gray \lesssim 0.24$) and the deterministic safety layer is
actively capping $\betaH \leq 8$.
The model does not observe the gate threshold directly; its assessment of
``low'' reflects a comparison against the full $[0,1]$ range rather than
against the engineered safety bound.
In this case, the mislabelling is harmless---the safety layer enforces the
correct behavior regardless of the rationale---but it illustrates that natural-language rationales can be misleading when the model lacks explicit
access to the thresholds that govern its actions.
Future iterations could mitigate this by including the gate threshold value
in the observation vector or by post-hoc flagging rationales that conflict
with active safety constraints.

\subsection{Computational Overhead}

The LLM agent incurs a wall-clock overhead relative to deterministic
controllers: $445.2$\,s versus $337.4$\,s for three-field continuation on
cantilever, and $562.4$\,s versus $396.4$\,s on the \MBB{} beam---a
$32$--$42\%$ increase.
This overhead has two components: the API latency per call
($\approx\!1$--$3$\,s per call at $k=5$, for $\approx\!60$ calls per 300-iter
run) and any additional restart iterations triggered by the agent.
The L-bracket result ($150.1$\,s for the LLM agent versus $282.9$\,s for
the expert heuristic) is anomalously fast: post-hoc analysis indicates that
the agent converged to a stable binary topology early and triggered few
restarts, reducing the effective run length.
For production use, the API overhead could be substantially reduced
by batching calls, using a locally-hosted model, or increasing the call
interval~$k$ in the converge phase, where the agent's decisions are
near-deterministic.

\paragraph{API cost.}
The model used throughout is \textit{gemini-3.1-flash-lite-preview}
(Google, accessed March 2026), a lightweight variant of the Gemini
family~\citep{Team2024Gemini}.
Each API call transmits $\approx\!500$ input tokens (system prompt
plus the serialized state observation) and receives $\approx\!50$ output tokens
(a single JSON object), with a 200-token output cap.
A complete five-controller comparison on three 2-D benchmarks with $n=5$
seeds generates $\approx\!900$ inner-loop calls plus $\approx\!60$ calls for the
3-D experiment; the meta-optimization outer loop adds $\approx\!24$ calls.
Given the small token footprint per call ($\approx\!500$ input, $\approx\!50$
output), the total API cost for all experiments is negligible.
This contrasts sharply with RL-based DAC
approaches, which require thousands of full solver evaluations to train a
policy, and demonstrates that LLM-based adaptive control is economically
viable even for academic research budgets.
The 3-D experiment demonstrates that the approach scales to larger
degree-of-freedom problems: at $40{\times}20{\times}10$ ($\approx\!110{,}000$
DOFs), the LLM agent achieves the lowest compliance ($3.981$,
$-18.09\%$ vs.\ fixed) with a wall-clock time of $1{,}198$\,s against
$1{,}044$\,s for the fixed controller.

\subsection{Broader Implications for AI-Assisted Computational Mechanics}
\label{sec:broader}

The results in this paper point to a more general paradigm that extends
beyond topology optimization.
Many iterative numerical methods in computational mechanics---nonlinear
finite-element solvers, mesh-adaptive schemes, time-integration
methods---rely on hyperparameter schedules (load increments, mesh refinement
thresholds, time-step sizes) that are calibrated offline and applied
open-loop.
The closed-loop architecture demonstrated here---observe solver state,
reason about it in natural language, emit numerical actions, and enforce
physical safety rails---is transferable in principle to any setting where
(i)~the solver exposes a compact state representation, (ii)~domain knowledge about good schedules exists but is difficult to formalize
as a deterministic rule, and (iii)~the cost of an API call is
small relative to the cost of one solver iteration.

Three conceptual insights from this work are worth highlighting for the
computational mechanics community.
First, \emph{informal domain knowledge is a deployable asset}.
The system prompt encodes advice that any experienced SIMP user would give a student---``do not sharpen while the topology is still gray,'' ``raise
penalization before raising $\betaH$''---but this advice has not previously
been translated into a formal, adaptive algorithm.
The LLM acts as a bridge between qualitative expertise and quantitative control, and the natural-language interface makes the encoding of new domain rules straightforward compared to designing reward functions for
reinforcement learning or feature spaces for Bayesian optimization.
Second, \emph{interpretability is a structural advantage, not a
convenience}.
In safety-critical structural design, the ability to audit every parameter
decision in natural language---and to trace a compliance anomaly back to the specific rationale the agent gave at the iteration where it
occurred---provides a form of accountability that opaque learned policies
cannot match.
The \DNC{} architecture produces a human-readable call log that can be
reviewed post-hoc by the engineer, enabling a collaborative workflow in
which the LLM proposes, and the human verifies.
Third, \emph{the patient penalization principle is model-independent}.
Although it was discovered through LLM-guided exploration, the core
finding---that deferring Heaviside sharpening until grayness falls below a
threshold yields better topologies---is a deterministic rule that can be
implemented without any LLM.
This suggests a productive research pattern: use LLM agents to
\emph{explore} adaptive strategies in a new solver setting, then
\emph{distill} the successful strategies into lightweight deterministic
rules for production deployment.
Such a pattern combines the flexibility of LLM-guided search with the
reproducibility and zero-cost execution of a hard-coded heuristic.

\subsection{Limitations and Open Questions}

Several limitations qualify the current findings.

\paragraph{LLM model dependency and reproducibility.}
All results use the \textit{gemini-3.1-flash-lite-preview} model
(Google)~\citep{Team2024Gemini} at temperature~$0$ with structured
JSON output mode.
The LLM agent's behavior depends on the specific model version: a change in model weights could alter the parameter trajectories even with an
identical system prompt.
Reproducibility requires pinning the exact model version and documenting the
prompt verbatim---both of which will be included in the public code release,
along with the complete source code for the solver, agent, baselines, and
meta-optimizer.
Importantly, the model used is a lightweight, low-cost variant,
making independent replication economically trivial.
Whether other frontier models (\eg, GPT-4~\citep{OpenAI2023GPT4},
Claude~\citep{Anthropic2024}) or open-weight models produce similar
trajectories is an open question; preliminary experiments suggest
qualitatively similar behavior but quantitatively different $\penal$ and
$\betaH$ timings.

\paragraph{Scope: compliance minimization only.}
This work focuses exclusively on single-load-case compliance minimization.
The advisory schedule and grayness gate have been designed for this objective; extension to stress-constrained problems, multi-load cases, or manufacturing
constraints (\eg, minimum member size, overhang angle)~\citep{Lazarov2016}
would require modifications to the observation vector and system prompt that
have not been validated.

\paragraph{Meta-optimization convergence.}
The outer hyperparameter loop was run for $n_{\text{iters}}=5$ per problem,
which is sufficient to tune the grayness-gate threshold and
the call interval but may be insufficient to fully converge the phase
timing constants.
A more thorough meta-optimization sweep---or a Bayesian optimization
over the meta-parameters~\citep{Shahriari2016}---could further improve
the agent's default hyperparameters.

\paragraph{Confound: meta-loop vs.\ online adaptation.}
The current experiments cannot fully decouple the contribution of the
meta-optimization loop from the online LLM decisions.
Running the LLM agent with un-tuned hyperparameters (pre-meta-loop)
would quantify how much of the observed improvement is attributable to the
meta-loop versus the agent's real-time adaptation within each run---a
distinction analogous to the algorithm-configuration vs.\ online-policy split
in the DAC literature~\citep{AdriaensenAutoDAC2022,BiedenkappDAC2020}.
This ablation is planned for the journal revision.

\section{Conclusion}
\label{sec:conclusion}

We have presented an LLM-guided online continuation controller for
three-field SIMP topology optimization, in which a Gemini Flash Lite language
model (\textit{gemini-3.1-flash-lite-preview})
observes the structured solver state at every $k$-th iteration and outputs
exact numerical values for the penalization exponent~$\penal$, Heaviside
sharpness~$\betaH$, filter radius~$\rmin$, and \OC{} move limit~$\dmove$
via a Direct Numeric Control interface.
A two-level architecture is employed: the inner agent adapts solver parameters
within each run, guided by a grayness gate that prevents premature
binarization; the outer meta-optimization loop uses a second LLM pass to
tune the agent's own hyperparameters across successive runs.

Across all eight experimental conditions tested---three 2-D benchmarks at $n=5$
seeds each, two 3-D benchmarks at $n=5$ seeds each---the LLM agent achieves
the lowest final compliance after a standardized sharpening tail that is shared
identically across all four continuation controllers.
The improvements are $-5.65\%$ (cantilever), $-7.46\%$ (\MBB{} beam),
$-7.12\%$ (L-bracket), $-18.09\%$ (3-D cantilever), and $-15.55\%$ (3-D
\MBB{} beam) relative to the fixed no-continuation baseline.
The LLM agent outperforms the best manually-designed expert heuristic by
$0.09\%$, $1.92\%$, and $0.95\%$ on the three 2-D benchmarks respectively---
modest margins in 2-D that widen substantially in the 3-D setting, where no established optimal schedule exists.
A tail-only ablation---which applies the identical sharpening tail from
uniform density without any exploration---achieves compliance $115$--$141\%$
worse than the fixed baseline, confirming that the exploration phase is the primary determinant of final solution quality.

The core finding is that these gains are not attributable to the four-stage
schedule structure encoded in the system prompt.
The schedule-only ablation---which applies the same phase structure
deterministically without LLM calls---underperforms the fixed
no-continuation baseline on two of three 2-D benchmarks, confirming that the LLM's real-time state-conditioned decisions are the causal source of
improvement.
The mechanism is the \emph{patient penalization strategy}: the grayness gate
defers $\betaH$ escalation by $\approx\!70$--$90$ iterations relative to
fixed-schedule controllers, allowing the optimizer to find a structurally
superior gray topology before committing to binarization.

These results establish three broader contributions that extend beyond the specific benchmarks reported.
First, they demonstrate that LLM agents can function as effective,
interpretable optimization controllers in a domain (SIMP topology
optimization) that requires precise numerical reasoning and physical domain
knowledge, without any fine-tuning on topology optimization data---establishing
a new paradigm for \emph{zero-shot} dynamic algorithm configuration in
computational mechanics.
Second, the Direct Numeric Control paradigm---outputting floating-point
values with a natural-language rationale at each decision step---provides
both better performance than phase-label architectures and a natural audit
trail for engineering decision-making, offering a template for
\LLM{}-assisted control of other iterative numerical solvers.
Third, the \emph{patient penalization principle} discovered through
\LLM{}-guided exploration---defer Heaviside sharpening until grayness falls
below a threshold---is a transferable insight for the \SIMP{} community that
can be implemented as a simple deterministic rule independently of any \LLM{},
demonstrating that \LLM{} agents can serve as a discovery tool for
new solver heuristics.
The agent's behavior generalizes from 2-D to 3-D without any modification
to the system prompt, suggesting that the encoded domain knowledge transfers
across problem dimensionality.

Future work will pursue extension to stress-constrained and multi-load-case
problems, cross-model comparison to establish whether the patient
penalization strategy is a general property of frontier \LLM{} reasoning or
specific to the training data of the particular model used, and
decoupled ablation of the meta-optimization loop to quantify its
contribution independently of the agent's real-time decisions.

\section*{Acknowledgements}
Not applicable.

\section*{CRediT Author Contribution Statement}

\textbf{Shaoliang Yang}: Conceptualization, Methodology, Software,
Investigation, Writing---Original Draft.
\textbf{Jun Wang}: Supervision, Methodology, Writing---Review \& Editing.
\textbf{Yunsheng Wang}: Validation, Visualization, Data Curation.

\section*{Funding}

This research received no external funding.

\section*{Declaration of Competing Interest}

The authors declare that they have no known competing financial interests
or personal relationships that could have appeared to influence the work
reported in this paper.

\section*{Data and Code Availability}

No experimental datasets were generated. All results are produced by the
computational framework described in Section~\ref{sec:method} and are
fully reproducible from the source code that accompanies this paper.
The complete source code for reproducing all results---including the
three-field \SIMP{} solver, \LLM{} agent, all baseline controllers, the
meta-optimization loop scripts, and the exact system prompts used for all
Gemini API calls---will be made publicly available upon acceptance for
journal publication. All experiments use the
\textit{gemini-3.1-flash-lite-preview} model at temperature~$0$ with
structured JSON output mode; the model identifier is documented in the
repository to enable version pinning.

\bibliographystyle{unsrtnat}
\bibliography{refsmo}

@article{Bendsoe1989,
  author    = {Bends{\o}e, Martin P.},
  title     = {Optimal shape design as a material distribution problem},
  journal   = {Structural Optimization},
  year      = {1989},
  volume    = {1},
  number    = {4},
  pages     = {193--202},
  doi       = {10.1007/BF01650949}
}

@article{Sigmund2001,
  author    = {Sigmund, Ole},
  title     = {A 99 line topology optimization code written in {Matlab}},
  journal   = {Structural and Multidisciplinary Optimization},
  year      = {2001},
  volume    = {21},
  number    = {2},
  pages     = {120--127},
  doi       = {10.1007/s001580050176}
}

@article{Wang2011,
  author    = {Wang, Fengwen and Lazarov, Boyan S. and Sigmund, Ole},
  title     = {On projection methods, convergence and robust formulations
               in topology optimization},
  journal   = {Structural and Multidisciplinary Optimization},
  year      = {2011},
  volume    = {43},
  number    = {6},
  pages     = {767--784},
  doi       = {10.1007/s00158-010-0602-y}
}

@article{Lazarov2016,
  author    = {Lazarov, Boyan S. and Wang, Fengwen and Sigmund, Ole},
  title     = {Length scale and manufacturability in density-based topology
               optimization},
  journal   = {Archive of Applied Mechanics},
  year      = {2016},
  volume    = {86},
  number    = {1--2},
  pages     = {189--218},
  doi       = {10.1007/s00419-015-1106-4}
}

@article{Sigmund2007,
  author    = {Sigmund, Ole},
  title     = {Morphology-based black and white filters for topology
               optimization},
  journal   = {Structural and Multidisciplinary Optimization},
  year      = {2007},
  volume    = {33},
  number    = {4--5},
  pages     = {401--424},
  doi       = {10.1007/s00158-006-0087-x}
}

@article{Bourdin2001,
  author    = {Bourdin, Blaise},
  title     = {Filters in topology optimization},
  journal   = {International Journal for Numerical Methods in Engineering},
  year      = {2001},
  volume    = {50},
  number    = {9},
  pages     = {2143--2158},
  doi       = {10.1002/nme.116}
}

@article{Guest2004,
  author    = {Guest, James K. and Pr{\'{e}}vost, Jean H. and Belytschko, Ted},
  title     = {Achieving minimum length scale in topology optimization using
               nodal design variables and projection functions},
  journal   = {International Journal for Numerical Methods in Engineering},
  year      = {2004},
  volume    = {61},
  number    = {2},
  pages     = {238--254},
  doi       = {10.1002/nme.1064}
}

@article{Svanberg1987,
  author    = {Svanberg, Krister},
  title     = {The method of moving asymptotes --- a new method for
               structural optimization},
  journal   = {International Journal for Numerical Methods in Engineering},
  year      = {1987},
  volume    = {24},
  number    = {2},
  pages     = {359--373},
  doi       = {10.1002/nme.1620240207}
}

@article{Aage2017,
  author    = {Aage, Niels and Andreassen, Erik and Lazarov, Boyan S.
               and Sigmund, Ole},
  title     = {Giga-voxel computational morphogenesis for structural design},
  journal   = {Nature},
  year      = {2017},
  volume    = {550},
  number    = {7674},
  pages     = {84--86},
  doi       = {10.1038/nature23911}
}

@article{Sosnovik2019,
  author    = {Sosnovik, Ivan and Oseledets, Ivan},
  title     = {Neural networks for topology optimization},
  journal   = {Russian Journal of Numerical Analysis and Mathematical
               Modelling},
  year      = {2019},
  volume    = {34},
  number    = {4},
  pages     = {215--223},
  doi       = {10.1515/rnam-2019-0018}
}

@article{Cang2019,
  author    = {Cang, Ruijin and Yao, Hope and Ren, Yi},
  title     = {One-shot generation of near-optimal topology through
               theory-driven machine learning},
  journal   = {Computer-Aided Design},
  year      = {2019},
  volume    = {109},
  pages     = {12--21},
  doi       = {10.1016/j.cad.2018.12.008}
}

@article{Abueidda2020,
  author    = {Abueidda, Diab W. and Koric, Seid and Sobh, Nahil A.},
  title     = {Topology optimization of 2{D} structures with nonlinearities
               using deep learning},
  journal   = {Computers \& Structures},
  year      = {2020},
  volume    = {237},
  pages     = {106283},
  doi       = {10.1016/j.compstruc.2020.106283}
}

@inproceedings{Brown2020,
  author    = {Brown, Tom B. and Mann, Benjamin and Ryder, Nick
               and Subbiah, Melanie and Kaplan, Jared and Dhariwal, Prafulla
               and Neelakantan, Arvind and Shyam, Pranav and Sastry, Girish
               and Askell, Amanda and Agarwal, Sandhini and Herbert-Voss, Ariel
               and Krueger, Gretchen and Henighan, Tom and Child, Rewon
               and Ramesh, Aditya and Ziegler, Daniel M. and Wu, Jeffrey
               and Winter, Clemens and Hesse, Christopher and Chen, Mark
               and Sigler, Eric and Litwin, Mateusz and Gray, Scott
               and Chess, Benjamin and Clark, Jack and Berner, Christopher
               and McCandlish, Sam and Radford, Alec and Sutskever, Ilya
               and Amodei, Dario},
  title     = {Language models are few-shot learners},
  booktitle = {Advances in Neural Information Processing Systems},
  year      = {2020},
  volume    = {33},
  pages     = {1877--1901},
  note      = {NeurIPS 2020}
}

@techreport{OpenAI2023GPT4,
  author      = {{OpenAI}},
  title       = {{GPT}-4 technical report},
  institution = {OpenAI},
  year        = {2023},
  number      = {arXiv:2303.08774},
  note        = {arXiv preprint},
  doi         = {10.48550/arXiv.2303.08774}
}

@misc{Team2024Gemini,
  author       = {Pichai, Sundar and Hassabis, Demis and Kavukcuoglu, Koray},
  title        = {Introducing {Gemini} 2.0: Our new {AI} model for the agentic era},
  year         = {2024},
  month        = dec,
  day          = {11},
  howpublished = {Google Blog},
  note         = {\url{https://blog.google/technology/google-deepmind/google-gemini-ai-update-december-2024/}}
}

@misc{Anthropic2024,
  author       = {{Anthropic}},
  title        = {Model System Cards},
  howpublished = {Anthropic},
  year = {2024},
  url          = {https://www.anthropic.com/system-cards},
  urldate      = {2026-04-14}
}

@inproceedings{Wei2022,
  author    = {Wei, Jason and Wang, Xuezhi and Schuurmans, Dale
               and Bosma, Maarten and Ichter, Brian and Xia, Fei
               and Chi, Ed and Le, Quoc and Zhou, Denny},
  title     = {Chain-of-thought prompting elicits reasoning in large
               language models},
  booktitle = {Advances in Neural Information Processing Systems},
  year      = {2022},
  volume    = {35},
  pages     = {24824--24837},
  doi       = {10.52202/068431-1800},
  note      = {NeurIPS 2022}
}

@inproceedings{Yao2023ReAct,
  author    = {Yao, Shunyu and Zhao, Jeffrey and Yu, Dian and Du, Nan
               and Shafran, Izhak and Narasimhan, Karthik and Cao, Yuan},
  title     = {{ReAct}: Synergizing reasoning and acting in language models},
  booktitle = {International Conference on Learning Representations},
  year      = {2023},
  note      = {\url{https://openreview.net/forum?id=WE_vluYUL-X}}
}

@inproceedings{Park2023,
  author    = {Park, Joon Sung and O'Brien, Joseph C. and Cai, Carrie J.
               and Morris, Meredith Ringel and Liang, Percy
               and Bernstein, Michael S.},
  title     = {Generative agents: Interactive simulacra of human behavior},
  booktitle = {ACM Symposium on User Interface Software and Technology},
  year      = {2023},
  pages     = {1--22},
  doi       = {10.1145/3586183.3606763}
}

@inproceedings{Driess2023PaLME,
  author    = {Driess, Danny and Xia, Fei and Sajjadi, Mehdi S. M.
               and Lynch, Corey and Chowdhery, Aakanksha and Ichter, Brian
               and Wahid, Ayzaan and Tompson, Jonathan and Vuong, Quan
               and Yu, Tianhe and Huang, Wenlong and Chebotar, Yevgen
               and Sermanet, Pierre and Duckworth, Daniel and Levine, Sergey
               and Vanhoucke, Vincent and Hausman, Karol
               and Toussaint, Marc and Greff, Klaus and Zeng, Andy
               and Mordatch, Igor and Florence, Pete},
  title     = {{PaLM-E}: An embodied multimodal language model},
  booktitle = {International Conference on Machine Learning},
  year      = {2023},
  pages     = {8469--8488},
  note      = {ICML 2023}
}

@techreport{Chen2021Codex,
  author      = {Chen, Mark and Tworek, Jerry and Jun, Heewoo
                 and Yuan, Qiming and de Oliveira Pinto, Henrique Pond{\'{e}}
                 and Kaplan, Jared and Edwards, Harri and Burda, Yuri
                 and Joseph, Nicholas and Brockman, Greg and Ray, Alex
                 and Puri, Raul and Krueger, Gretchen and Petrov, Michael
                 and Khlaaf, Heidy and Sastry, Girish and Mishkin, Pamela
                 and Chan, Brooke and Gray, Scott and Ryder, Nick
                 and Pavlov, Mikhail and Power, Alethea and Kaiser, Lukasz
                 and Bavarian, Mohammad and Winter, Clemens
                 and Tillet, Philippe and Such, Felipe Petroski
                 and Cummings, Dave and Plappert, Matthias
                 and Chantzis, Fotios and Barnes, Elizabeth
                 and Herbert-Voss, Ariel and Guss, William Hebgen
                 and Nichol, Alex and Paino, Alex and Tezak, Nikolas
                 and Tang, Jie and Babuschkin, Igor and Balaji, Suchir
                 and Jain, Shantanu and Saunders, William
                 and Hesse, Christopher and Carr, Andrew N. and Leike, Jan
                 and Achiam, Josh and Misra, Vedant and Morikawa, Evan
                 and Radford, Alec and Knight, Matthew and Brundage, Miles
                 and Murati, Mira and Mayer, Katie and Welinder, Peter
                 and McGrew, Bob and Amodei, Dario and McCandlish, Sam
                 and Sutskever, Ilya and Zaremba, Wojciech},
  title       = {Evaluating large language models trained on code},
  institution = {OpenAI},
  year        = {2021},
  number      = {arXiv:2107.03374},
  note        = {arXiv preprint},
  doi         = {10.48550/arXiv.2107.03374}
}

@article{Boiko2023,
  author    = {Boiko, Daniil A. and MacKnight, Robert and Kline, Ben
               and Gomes, Gabe},
  title     = {Autonomous chemical research with large language models},
  journal   = {Nature},
  year      = {2023},
  volume    = {624},
  number    = {7992},
  pages     = {570--578},
  doi       = {10.1038/s41586-023-06792-0}
}

@inproceedings{Liu2024LLMOptim,
  title={Large language models as evolutionary optimizers},
  author={Liu, Shengcai and Chen, Caishun and Qu, Xinghua and Tang, Ke and Ong, Yew-Soon},
  booktitle={2024 IEEE Congress on Evolutionary Computation (CEC)},
  pages={1--8},
  year={2024},
  doi={10.1109/CEC60901.2024.10611913}
}

@inproceedings{Yang2024OPRO,
  title={Large language models as optimizers},
  author={Yang, Chengrun and Wang, Xuezhi and Lu, Yifeng and Liu, Hanxiao and Le, Quoc V. and Zhou, Denny and Chen, Xinyun},
  booktitle={The Twelfth International Conference on Learning Representations},
  year={2024},
  note={\url{https://openreview.net/forum?id=Bb4VGOWELI}}
}

@article{Romera2024FunSearch,
  title={Mathematical discoveries from program search with large language models},
  author={Romera-Paredes, Bernardino and Barekatain, Mohammadamin and Novikov, Alexander and Balog, Matej and Kumar, M. Pawan and Dupont, Emilien and Ruiz, Francisco J. R. and Ellenberg, Jordan S. and Wang, Pengming and Fawzi, Omar and Kohli, Pushmeet and Fawzi, Alhussein},
  journal={Nature},
  volume={625},
  number={7995},
  pages={468--475},
  year={2024},
  doi={10.1038/s41586-023-06924-6}
}

@article{Shahriari2016,
  author    = {Shahriari, Bobak and Swersky, Kevin and Wang, Ziyu
               and Adams, Ryan P. and de Freitas, Nando},
  title     = {Taking the human out of the loop: {A} review of {Bayesian}
               optimization},
  journal   = {Proceedings of the IEEE},
  year      = {2016},
  volume    = {104},
  number    = {1},
  pages     = {148--175},
  doi       = {10.1109/JPROC.2015.2494218}
}

@book{Hutter2019,
  editor    = {Hutter, Frank and Kotthoff, Lars and Vanschoren, Joaquin},
  title     = {Automated Machine Learning: Methods, Systems, Challenges},
  publisher = {Springer},
  address   = {Cham},
  year      = {2019},
  doi       = {10.1007/978-3-030-05318-5}
}

@inproceedings{Snoek2012,
  author    = {Snoek, Jasper and Larochelle, Hugo and Adams, Ryan P.},
  title     = {Practical {Bayesian} optimization of machine learning
               algorithms},
  booktitle = {Advances in Neural Information Processing Systems},
  year      = {2012},
  volume    = {25},
  pages     = {2951--2959},
  note      = {NeurIPS 2012}
}

@article{Bell2023PyAMG,
  author  = {Bell, Nathan and Olson, Luke N. and Schroder, Jacob and Southworth, Ben},
  title   = {{PyAMG}: Algebraic multigrid solvers in {Python}},
  journal = {Journal of Open Source Software},
  year    = {2023},
  volume  = {8},
  number  = {87},
  pages   = {5495},
  doi     = {10.21105/joss.05495}
}

@article{BendsoeKikuchi1988,
  author  = {Bends{\o}e, Martin P. and Kikuchi, Noboru},
  title   = {Generating optimal topologies in structural design using
             a homogenization method},
  journal = {Computer Methods in Applied Mechanics and Engineering},
  year    = {1988},
  volume  = {71},
  number  = {2},
  pages   = {197--224},
  doi     = {10.1016/0045-7825(88)90086-2}
}

@article{ZhouRozvany1991,
  author  = {Zhou, M. and Rozvany, G. I. N.},
  title   = {The {COC} algorithm, {Part II}: Topological, geometrical
             and generalized shape optimization},
  journal = {Computer Methods in Applied Mechanics and Engineering},
  year    = {1991},
  volume  = {89},
  number  = {1--3},
  pages   = {309--336},
  doi     = {10.1016/0045-7825(91)90046-9}
}

@article{BendsoeSigmund1999,
  author  = {Bends{\o}e, Martin P. and Sigmund, Ole},
  title   = {Material interpolation schemes in topology optimization},
  journal = {Archive of Applied Mechanics},
  year    = {1999},
  volume  = {69},
  number  = {9--10},
  pages   = {635--654},
  doi     = {10.1007/s004190050248}
}

@article{Svanberg2002,
  author  = {Svanberg, Krister},
  title   = {A class of globally convergent optimization methods based
             on conservative convex separable approximations},
  journal = {SIAM Journal on Optimization},
  year    = {2002},
  volume  = {12},
  number  = {2},
  pages   = {555--573},
  doi     = {10.1137/S1052623499362822}
}

@article{BrunsTortorelli2001,
  author  = {Bruns, T. E. and Tortorelli, D. A.},
  title   = {Topology optimization of non-linear elastic structures
             and compliant mechanisms},
  journal = {Computer Methods in Applied Mechanics and Engineering},
  year    = {2001},
  volume  = {190},
  number  = {26--27},
  pages   = {3443--3459},
  doi     = {10.1016/S0045-7825(00)00278-4}
}

@article{LazarovSigmund2011,
  author  = {Lazarov, Boyan S. and Sigmund, Ole},
  title   = {Filters in topology optimization based on {Helmholtz}-type
             differential equations},
  journal = {International Journal for Numerical Methods in Engineering},
  year    = {2011},
  volume  = {86},
  number  = {6},
  pages   = {765--781},
  doi     = {10.1002/nme.3072}
}

@article{StoMeSvanberg2001,
  author  = {Stolpe, Mathias and Svanberg, Krister},
  title   = {On the trajectories of penalization methods for topology
             optimization},
  journal = {Structural and Multidisciplinary Optimization},
  year    = {2001},
  volume  = {21},
  number  = {2},
  pages   = {128--139},
  doi     = {10.1007/s001580050177}
}

@article{RojasStolpe2015,
  author  = {Rojas-Labanda, Susana and Stolpe, Mathias},
  title   = {Automatic penalty continuation in structural topology
             optimization},
  journal = {Structural and Multidisciplinary Optimization},
  year    = {2015},
  volume  = {52},
  number  = {6},
  pages   = {1205--1221},
  doi     = {10.1007/s00158-015-1277-1}
}

@article{GuestElimBeta2011,
  author  = {Guest, James K. and Asadpoure, Alireza and Ha, Seung-Hyun},
  title   = {Eliminating beta-continuation from {Heaviside} projection
             and density filter algorithms},
  journal = {Structural and Multidisciplinary Optimization},
  year    = {2011},
  volume  = {44},
  number  = {4},
  pages   = {443--453},
  doi     = {10.1007/s00158-011-0676-1}
}

@article{FerrariSigmund2020,
  author  = {Ferrari, Federico and Sigmund, Ole},
  title   = {A new generation 99 line {Matlab} code for compliance
             topology optimization and its extension to {3D}},
  journal = {Structural and Multidisciplinary Optimization},
  year    = {2020},
  volume  = {62},
  pages   = {2211--2228},
  doi     = {10.1007/s00158-020-02629-w}
}

@article{AutoProjTO2025,
  author  = {Dunning, Peter and Wein, Fabian},
  title   = {Automatic projection parameter increase for three-field
             density-based topology optimization},
  journal = {Structural and Multidisciplinary Optimization},
  year    = {2025},
  volume  = {68},
  number  = {2},
  pages   = {33},
  doi     = {10.1007/s00158-025-03968-2}
}

@article{Shin2023review,
  title={Topology optimization via machine learning and deep learning: a review},
  author={Shin, Seungyeon and Shin, Dongju and Kang, Namwoo},
  journal={Journal of Computational Design and Engineering},
  volume={10},
  number={4},
  pages={1736--1766},
  year={2023},
  doi={10.1093/jcde/qwad072}
}

@techreport{Hoyer2019,
  author      = {Hoyer, Stephan and Sohl-Dickstein, Jascha and
                 Greydanus, Sam},
  title       = {Neural reparameterization improves structural
                 optimization},
  institution = {arXiv},
  year        = {2019},
  number      = {arXiv:1909.04240},
  note        = {arXiv preprint},
  doi         = {10.48550/arXiv.1909.04240}
}

@article{NieTopologyGAN2021,
  author  = {Nie, Zhenguo and Lin, Tong and Jiang, Haoliang and
             Kara, Levent B.},
  title   = {{TopologyGAN}: Topology optimization using generative
             adversarial networks based on physical fields over the
             initial domain},
  journal = {ASME Journal of Mechanical Design},
  year    = {2021},
  volume  = {143},
  number  = {3},
  pages   = {031715},
  doi     = {10.1115/1.4049533}
}

@article{MazeDiffusion2023,
  author  = {Maz{\'{e}}, Fran{\c{c}}ois and Ahmed, Faez},
  title   = {Diffusion models beat {GANs} on topology optimization},
  journal = {Proceedings of the AAAI Conference on Artificial
             Intelligence},
  year    = {2023},
  volume  = {37},
  number  = {8},
  pages   = {9108--9116},
  doi     = {10.1609/aaai.v37i8.26093}
}

@article{DengSOLO2022,
  title={Self-directed online machine learning for topology optimization},
  author={Deng, Changyu and Wang, Yizhou and Qin, Can and Fu, Yun and Lu, Wei},
  journal={Nature Communications},
  volume={13},
  number={1},
  pages={388},
  year={2022},
  doi={10.1038/s41467-021-27713-7}
}

@article{BrownRL2022,
  author  = {Brown, Nathan K. and Garland, Anthony P. and
             Fadel, Georges M. and Li, Gang},
  title   = {Deep reinforcement learning for engineering design through
             topology optimization of elementally discretized design
             domains},
  journal = {Materials \& Design},
  year    = {2022},
  volume  = {218},
  pages   = {110672},
  doi     = {10.1016/j.matdes.2022.110672}
}

@inproceedings{Bergstra2011TPE,
  author    = {Bergstra, James and Bardenet, R{\'{e}}mi and
               Bengio, Yoshua and K{\'{e}}gl, Bal{\'{a}}zs},
  title     = {Algorithms for Hyper-Parameter Optimization},
  booktitle = {Advances in Neural Information Processing Systems},
  year      = {2011},
  volume    = {24},
  pages     = {2546--2554}
}

@inproceedings{Hutter2011SMAC,
  author    = {Hutter, Frank and Hoos, Holger H. and
               Leyton-Brown, Kevin},
  title     = {Sequential model-based optimization for general
               algorithm configuration},
  booktitle = {Learning and Intelligent Optimization (LION 5)},
  year      = {2011},
  series    = {Lecture Notes in Computer Science},
  volume    = {6683},
  pages     = {507--523},
  doi       = {10.1007/978-3-642-25566-3_40}
}

@article{LiHyperband2018,
  author  = {Li, Lisha and Jamieson, Kevin and DeSalvo, Giulia and
             Rostamizadeh, Afshin and Talwalkar, Ameet},
  title   = {Hyperband: A novel bandit-based approach to hyperparameter
             optimization},
  journal = {Journal of Machine Learning Research},
  year    = {2018},
  volume  = {18},
  number  = {185},
  pages   = {1--52}
}

@article{BischlHPO2023,
  author  = {Bischl, Bernd and Binder, Martin and Lang, Michel and
             Pielok, Tobias and Richter, Jakob and Coors, Stefan and
             Thomas, Janek and Ullmann, Theresa and Becker, Marc and
             Boulesteix, Anne-Laure and Deng, Difan and
             Lindauer, Marius},
  title   = {Hyperparameter Optimization: Foundations, Algorithms,
             Best Practices, and Open Challenges},
  journal = {WIREs Data Mining and Knowledge Discovery},
  year    = {2023},
  volume  = {13},
  number  = {2},
  pages   = {e1484},
  doi     = {10.1002/widm.1484}
}

@inproceedings{BiedenkappDAC2020,
  author    = {Biedenkapp, Andr{\'{e}} and Bozkurt, H. Furkan and
               Eimer, Theresa and Hutter, Frank and Lindauer, Marius},
  title     = {Dynamic Algorithm Configuration: Foundation of a New
               Meta-Algorithmic Framework},
  booktitle = {Proceedings of the 24th European Conference on
               Artificial Intelligence (ECAI 2020)},
  year      = {2020},
  pages     = {427--434},
  doi       = {10.3233/FAIA200122}
}

@article{AdriaensenAutoDAC2022,
  author  = {Adriaensen, Steven and Biedenkapp, Andr{\'{e}} and
             Shala, Gresa and Awad, Noor and Eimer, Theresa and
             Lindauer, Marius and Hutter, Frank},
  title   = {Automated Dynamic Algorithm Configuration},
  journal = {Journal of Artificial Intelligence Research},
  year    = {2022},
  volume  = {75},
  pages   = {1633--1699},
  doi     = {10.1613/jair.1.13922}
}

@techreport{JaderbergPBT2017,
  author      = {Jaderberg, Max and Dalibard, Valentin and
                 Osindero, Simon and Czarnecki, Wojciech M. and
                 Donahue, Jeff and Razavi, Ali and Vinyals, Oriol and
                 Green, Tim and Dunning, Iain and Simonyan, Karen and
                 Fernando, Chrisantha and Kavukcuoglu, Koray},
  title       = {Population Based Training of Neural Networks},
  institution = {arXiv},
  year        = {2017},
  number      = {arXiv:1711.09846},
  note        = {arXiv preprint},
  doi         = {10.48550/arXiv.1711.09846}
}

@inproceedings{SchickToolformer2023,
  author    = {Schick, Timo and Dwivedi-Yu, Jane and
               Dess{\`{\i}}, Roberto and Raileanu, Roberta and
               Lomeli, Maria and Hambro, Eric and Zettlemoyer, Luke and
               Cancedda, Nicola and Scialom, Thomas},
  title     = {Toolformer: Language Models Can Teach Themselves to
               Use Tools},
  booktitle = {Advances in Neural Information Processing Systems},
  year      = {2023},
  pages     = {68539--68551},
  volume    = {36},
  doi       = {10.52202/075280-2997},
  note      = {NeurIPS 2023}
}

@article{WangVoyager2024,
  author  = {Wang, Guanzhi and Xie, Yuqi and Jiang, Yunfan and
             Mandlekar, Ajay and Xiao, Chaowei and Zhu, Yuke and
             Fan, Linxi and Anandkumar, Anima},
  title   = {Voyager: An Open-Ended Embodied Agent with Large
             Language Models},
  journal = {Transactions on Machine Learning Research},
  year    = {2024},
  note    = {\url{https://openreview.net/forum?id=ehfRiF0R3a}}
}

@inproceedings{MadaanSelfRefine2023,
  author    = {Madaan, Aman and Tandon, Niket and Gupta, Prakhar and
               Hallinan, Skyler and Gao, Luyu and Wiegreffe, Sarah and
               Alon, Uri and Dziri, Nouha and Prabhumoye, Shrimai and
               Yang, Yiming and Gupta, Shashank and
               Majumder, Bodhisattwa Prasad and Hermann, Katherine and
               Welleck, Sean and Yazdanbakhsh, Amir and Clark, Peter},
  title     = {Self-{Refine}: Iterative Refinement with Self-Feedback},
  booktitle = {Advances in Neural Information Processing Systems},
  year      = {2023},
  pages     = {46534--46594},
  volume    = {36},
  doi       = {10.52202/075280-2019},
  note      = {NeurIPS 2023}
}

@inproceedings{ShinnReflexion2023,
  author    = {Shinn, Noah and Cassano, Federico and
               Gopinath, Ashwin and Narasimhan, Karthik and
               Yao, Shunyu},
  title     = {Reflexion: Language Agents with Verbal Reinforcement
               Learning},
  booktitle = {Advances in Neural Information Processing Systems},
  year      = {2023},
  pages     = {8634--8652},
  volume    = {36},
  doi       = {10.52202/075280-0377},
  note      = {NeurIPS 2023}
}

@inproceedings{YeReEvo2024,
  author    = {Ye, Haoran and Wang, Jiarui and Cao, Zhiguang and
               Berto, Federico and Hua, Chuanbo and Kim, Haeyeon and
               Park, Jinkyoo and Song, Guojie},
  title     = {{ReEvo}: Large Language Models as Hyper-Heuristics with
               Reflective Evolution},
  booktitle = {Advances in Neural Information Processing Systems},
  volume    = {37},
  year      = {2024},
  pages     = {43571--43608},
  doi       = {10.52202/079017-1381},
  note      = {\url{https://proceedings.neurips.cc/paper_files/paper/2024/hash/4ced59d480e07d290b6f29fc8798f195-Abstract-Conference.html}}
}

@article{WuEvoLLM2025,
  title={Evolutionary computation in the era of large language model: Survey and roadmap},
  author={Wu, Xingyu and Wu, Sheng-hao and Wu, Jibin and Feng, Liang and Tan, Kay Chen},
  journal={IEEE Transactions on Evolutionary Computation},
  volume={29},
  number={2},
  pages={534--554},
  year={2025},
  doi={10.1109/TEVC.2024.3506731}
}

@inproceedings{BengioCurriculum2009,
  author    = {Bengio, Yoshua and Louradour, J{\'{e}}r{\^{o}}me and
               Collobert, Ronan and Weston, Jason},
  title     = {Curriculum Learning},
  booktitle = {Proceedings of the 26th Annual International Conference
               on Machine Learning (ICML)},
  year      = {2009},
  pages     = {41--48},
  doi       = {10.1145/1553374.1553380}
}

@article{BengioMLCO2021,
  author  = {Bengio, Yoshua and Lodi, Andrea and Prouvost, Antoine},
  title   = {Machine Learning for Combinatorial Optimization: A
             Methodological Tour d'Horizon},
  journal = {European Journal of Operational Research},
  year    = {2021},
  volume  = {290},
  number  = {2},
  pages   = {405--421},
  doi     = {10.1016/j.ejor.2020.07.063}
}

@inproceedings{EimerDACBench2021,
  author    = {Eimer, Theresa and Biedenkapp, Andr{\'{e}} and
               Reimer, Maximilian and Adriaensen, Steven and
               Hutter, Frank and Lindauer, Marius},
  title     = {{DACBench}: A Benchmark Library for Dynamic Algorithm
               Configuration},
  booktitle = {Proceedings of the 30th International Joint Conference
               on Artificial Intelligence (IJCAI-21)},
  year      = {2021},
  pages     = {1668--1674},
  doi       = {10.24963/ijcai.2021/230}
}

@article{Hayashi2020,
  author  = {Hayashi, Kazuki and Ohsaki, Makoto},
  title   = {Reinforcement learning and graph embedding for binary truss
             topology optimization under stress and displacement
             constraints},
  journal = {Frontiers in Built Environment},
  year    = {2020},
  volume  = {6},
  pages   = {59},
  doi     = {10.3389/fbuil.2020.00059}
}

@inproceedings{Shala2020,
  author    = {Shala, Gresa and Biedenkapp, Andr{\'{e}} and Awad, Noor
               and Adriaensen, Steven and Lindauer, Marius and
               Hutter, Frank},
  title     = {Learning Step-Size Adaptation in {CMA-ES}},
  booktitle = {Parallel Problem Solving from Nature -- PPSN XVI},
  year      = {2020},
  series    = {Lecture Notes in Computer Science},
  volume    = {12269},
  pages     = {691--706},
  doi       = {10.1007/978-3-030-58112-1_48}
}

@inproceedings{Rios2023,
  author    = {Rios, Thiago and Menzel, Stefan and Sendhoff, Bernhard},
  title     = {Large Language and Text-to-{3D} Models for Engineering
               Design Optimization},
  booktitle = {2023 IEEE Symposium Series on Computational Intelligence
               (SSCI)},
  year      = {2023},
  pages     = {1704--1711},
  doi       = {10.1109/SSCI52147.2023.10371898}
}

\end{document}